\documentclass[a4paper,aps,prd,showpacs,nofootinbib,preprintnumbers,amsmath,amssymb,mcite,superscriptaddress,12pt]{revtex4-1}
\usepackage[table]{xcolor}
\usepackage{graphicx}
\usepackage{subfigure}
\usepackage{dcolumn}
\usepackage{bm}
\usepackage{esvect}
\usepackage{accents}
\usepackage{slashed}

\usepackage{hyperref}
\hypersetup{colorlinks=true,linkcolor=magenta,anchorcolor=green,citecolor=cyan,filecolor=black,menucolor=black,urlcolor=brown}
\usepackage{slashed}

\newcommand{\be}{\begin{equation}}
\newcommand{\ee}{\end{equation}}
\newcommand{\ba}{\begin{eqnarray}}
\newcommand{\ea}{\end{eqnarray}}

\newcommand{\Mpl}{M_{\textrm{Pl}}}

\begin{document}

\title{$(g-2)_\mu$ and Screened Modified Gravity}

\author{Philippe Brax}
\affiliation{Institut de Physique Th\'eorique, Universit\'e  Paris-Saclay, CEA, CNRS, F-91191 Gif-sur-Yvette Cedex, France}
\author{Anne-Christine Davis}
\affiliation{DAMTP, Centre for Mathematical Sciences, University of Cambridge, Wilberforce Road, Cambridge CB3 0WA, United Kingdom}
\author{Benjamin Elder}
\affiliation{Department of Physics and Astronomy, University of Hawai'i, 2505 Correa Road, Honolulu, HI 96822, USA}
\begin{abstract} \footnotesize
We show how light scalar fields could account for the discrepancy between the theoretical and observed values of the anomalous magnetic moment of the (anti)muon. When coupled  to both matter and photons,  light scalar fields  induce a change of the anomalous magnetic moment of charged particles. This arises from two concurrent effects. Classically, light scalars induce a change of the cyclotron frequency, complementing the electromagnetic effects coming from the magnetic and electric fields used experimentally. Light scalars also contribute to the anomalous magnetic moment quantum mechanically at the one-loop level. For unscreened scalar fields coupling with a Yukawa interaction to matter, these contributions are negligible after applying the Cassini bound on deviations from Newtonian gravity. On the other hand, screened scalars such as chameleons or symmetrons can couple strongly to matter in the laboratory and decouple in the Solar System. This allows us to probe branches of their parameter spaces where the recently measured anomalous magnetic moment of the (anti)muon can be accounted for in the chameleon and symmetron cases. We consider the compatibility of these models with other  cosmological and particle physics observables. We find that prototype chameleon and symmetron models considered here are in tension with the bounds on the branching ratio of  kaons into pions and invisible matter.

\end{abstract}

\maketitle

\section{Introduction}

The recent measurement of the anomalous magnetic moment of (anti)muons  deviates at the 4.2-$\sigma$ level from the expected value in the Standard Model~\cite{Muong-2:2021ojo}.  This result aligns with an earlier measurement by BNL~\cite{Muong-2:2006rrc}, motivating theory explanations for this anomaly. The Standard Model prediction was discussed by the international theory collaboration (see \cite{Aoyama:2020ynm}, as well as ~\cite{Aoyama:2012wk,Aoyama:2019ryr,Czarnecki:2002nt,Gnendiger:2013pva,Davier:2017zfy,Keshavarzi:2018mgv,Colangelo:2018mtw,Hoferichter:2019mqg,Davier:2019can,Keshavarzi:2019abf,Kurz:2014wya,Melnikov:2003xd,Masjuan:2017tvw,Colangelo:2017fiz,Hoferichter:2018kwz,Gerardin:2019vio,Bijnens:2019ghy,Colangelo:2019uex,Blum:2019ugy,Colangelo:2014qya}).\footnote{An alternative computation has been presented in~\cite{Borsanyi:2020mff} which removes the muon $g - 2$ discrepancy, but it does so by modifying the hadronic vacuum polarization in a way that raises tensions elsewhere~\cite{Crivellin:2020zul,Keshavarzi:2020bfy,Colangelo:2020lcg}.} Most explanations of this result involve specific extensions of the Standard Model such as supersymmetric physics or axions. The corrections induced by the Beyond the Standard Model (BSM) models appear at the quantum level and complement the contributions coming from Standard Model radiative corrections, for example see~\cite{Buttazzo:2020vfs,Baum:2021qzx,Buen-Abad:2021fwq,Brdar:2021pla,Ge:2021cjz}. In this paper we show there is another, hitherto unexplored, candidate to explain this discrepancy: a light scalar field coupled to matter.

Scalar fields are a generic prediction of modifications to General Relativity, and were originally motivated by the late time acceleration of the Universe. In this context, we specifically concentrate on light scalar fields coupled to matter. Of course, these fields mediate a ``fifth force'' and therefore there are strong bounds by laboratory~\cite{Adelberger:2003zx} and Solar System experiments~\cite{Will:1993ns} such as the Cassini~\cite{Bertotti:2003rm} probe which give tight constraints on the coupling strength with atoms.  Because of this, we focus on models where {\it screening} takes place, that is, a dynamical suppression of the fifth force due to environmental interactions.  This scenario was extensively reviewed in \cite{Joyce:2014kja}.  We will assume that the coupling to matter is universal and in particular couples to all fermions of the Standard Model in the same way;
we will focus on the couplings to the leptons. 

Such light scalars induce two types of corrections to the anomalous magnetic moment of leptons. The first one is an experimental effect, i.e. it enters only in the comparison with experimental results  as the classical effects of light scalars on the cyclotron frequency  are not taken into account in the data analysis. Their evaluation requires a careful reexamination of the anomalous spin precession  usually deduced from the Bargmann-Michel-Telegdi (BMT) equation~\cite{Bargmann:1959gz}, see also~\cite{Jackson:1998nia} for a derivation which takes into account  electromagnetic effects only. The second contribution to the anomalous magnetic momentum of leptons comes from the one loop diagrams including the propagation of light scalars in the loops. This is well-known and needs to be added to the classical effect before comparing to the recent experimental results. 

Here we concentrate on two typical models in which screening takes place. They are also archetypical examples of two screening mechanisms. The first one is the chameleon mechanism~\cite{Khoury:2003aq} whereby the scalar mass increases with the ambient density of the environment.  The consequences of this are twofold: first, the scalar charge carried by dense, macroscopic objects is exponentially suppressed.  Second, inside a comparatively dense environment like the Earth's atmosphere the fifth force is very short-ranged. The second screening mechanism we consider is the Damour-Polyakov mechanism~\cite{Damour:1994zq, Olive:2007aj,Hinterbichler:2010es} whereby the coupling strength to  matter is depleted in dense matter. In both cases, the respective screening mechanisms allow for strong fifth force interactions in vacuum and highly suppressed ones in dense environments. The original inverse power law chameleon, referred to as the  chameleon in the following, will be used for the chameleon mechanism. The symmetron~\cite{Hinterbichler:2010es} will be used for the Damour-Polyakov mechanism~\cite{Damour:1994zq}.  
The models are strongly constrained by laboratory experiments such as atom interferometry~\cite{Hamilton:2015zga, Elder:2016yxm, Burrage:2016rkv, Jaffe:2016fsh, Sabulsky:2018jma}, the Casimir effect~\cite{Brax:2007vm, Elder:2019yyp} or neutron bouncers~\cite{Jenke:2014yel, Jenke:2020obe}. For a review of these tests, as well as astrophysical bounds, see~\cite{Burrage:2016bwy, CANTATA:2021ktz}. We will find that neither the classical chameleon nor the classical symmetron can account for the anomalous muon's results. However, in both cases once quantum corrections are taken into account there is a region of parameter space that could account for the muon results, thanks to the large couplings that are enabled by the screening mechanisms.  This is a regime where the quantum corrections dominate over the classical effects. 


The plan of this paper is as follows. In Section 2 we introduce the screened modified gravity models under consideration. In Section 3 we explain how the anomalous magnetic moment of a particle like the muon, or electron, is determined by comparing the cyclotron frequency to the spin precession frequency. We include the effect of the scalar particle in the derivation and give a complete derivation of the anomalous spin precession.
This section shows how the scalar particle changes the classical result for the anomalous magnetic moment. 
Section 4 contains a derivation of the quantum corrections for the scalar particle. We then apply this to the anomalous magnetic moment of the muon in section 5, first showing that a standard Yukawa interaction could not account for the discrepancy due to the strict Solar System constraints. We show that both chameleon models and symmetrons could account for the anomalous magnetic moment of the (anti) muon as measured by Fermilab and BNL before~\cite{Muong-2:2006rrc}. This result requires a careful investigation of the parameter space of both models against other experiments and, in the case of the chameleon, recalculating the bounds from the Hydrogen transition taking screening into account. In Section 6 we discuss the compatibility of our parameters with other constraints on the two models coming from cosmology, such as BBN and the CMB, and from LHC physics such as the Higgs decay. Here we find that most of the chameleon and symmetron  models which are compatible with $(g-2)_\mu$ have hardly any cosmological impact. On the other hand, we show that the scalars must be essentially decoupled from the Higgs boson for the screening behavior of the theories to be preserved. We also find that the explanation of $(g-2)_\mu$ by chameleons and symmetrons is in tension with results from  kaon decays into pions and invisible matter. This tension could be relaxed if  the universality of the coupling to matter  is broken.
We conclude in Section 7. There are three technical appendices where we discuss the angular momentum tensor, give a simplified derivation of the shift in the classical frequency, 
and show that our models are modified gravity, rather than dark energy.

\section{Screened modified gravity}
\subsection{The models}
We begin this section by briefly reviewing the ingredients that constitute chameleon and symmetron models. Both belong to the same family of scalar field theories governed by the Lagrangian\footnote{For the purposes of laboratory experiments, it suffices to work in flat space. See, e.g., the reviews in Refs.~\cite{Clifton:2011jh,Joyce:2014kja} for the covariant form of this action. Our metric signature is $(-,+,+,+)$, we work in units where $c = \hbar = 1$, and we define the reduced Planck mass as $m_{\rm pl} = (8 \pi G)^{-1/2}$~.}
\begin{equation}
\label{eq:L}
\mathcal L = -\frac{1}{2}(\partial\phi)^2 - V(\phi) + \mathcal L_m(\Psi,A^2(\phi) \eta_{\mu\nu})~,
\end{equation}
where the Standard Model fields (denoted collectively by $\Psi$) and their couplings to $\phi$ are encapsulated in the third term $\mathcal L_m$. This dependence appears in the minimal coupling to the rescaled metric $g_{\mu\nu}=A^2(\phi) \eta_{\mu\nu}$.
In these models the coupling to leptons is given by
\be 
\beta (\phi)= m_{\rm Pl} \frac{\partial \ln A}{\partial \phi}~.
\ee
Given the two functions $A(\phi)$ and $V(\phi)$, the scalar dynamics are governed by the Klein-Gordon equation
\be 
\Box \phi + \frac{\partial V_{\rm eff}}{\partial \phi}=0~.
\ee
The effective potential is given by
\be 
V_{\rm eff}(\phi)= V(\phi) + (A(\phi)-1) \rho_m~,
\ee
where $\rho_m$ is the conserved matter density. 
In homogeneous matter, the field settles at the minimum $\phi(\rho_m)$ of the effective potential. The scalar field's mass is given by
\be 
m^2(\rho_m)= \frac{\partial ^2 V_\mathrm{eff}}{\partial \phi^2}\bigg\vert_{\phi(\rho_m)}~,
\ee
and its effective coupling is
\be 
\beta (\rho_m)= \beta (\phi(\rho_m))~.
\ee
Roughly speaking, one expects chameleon screening in a body of size $R$ and density $\rho_m$ when $m(\rho_m) R\gg 1$ and Damour-Polyakov screening when $\beta (\rho_m)=0$.

\subsection{Chameleon and symmetron}

We will focus on chameleon models with
 the potential and matter coupling
\begin{equation}
     \quad \quad V(\phi) = \Lambda^4 \left( 1 + \frac{\Lambda^n}{\phi^n} \right)~, \quad \quad A(\phi) = 1 + \frac{\phi}{M}~.
     \label{chameleon-potentials}
\end{equation}
and likewise symmetrons have
\begin{equation}
    V(\phi) = - \frac{1}{2} \mu^2 \phi^2 + \frac{\lambda}{4} \phi^4~, \quad \quad A(\phi) = 1 + \frac{\phi^2}{2 M^2}~.
    \label{symmetron-potentials}
\end{equation}
In the symmetron case we have
\be 
\beta(\phi)\sim  \frac{m_{\rm Pl}\phi}{M^2}~,
\ee
which vanishes in very dense environment where $\phi$ settles at the origin.
The mass of the symmetron is of order $\mu$ which must be close to the inverse size of the experiments to be probed for there to be an appreciable classical effect.  If it is much larger the scalar field is exponentially suppressed thanks to its large mass, and if it is much smaller then the field does not roll away from $\phi = 0$~\cite{Upadhye:2012rc}. The latest experimental constraints on symmetrons can be found in \cite{Jaffe:2016fsh, Sabulsky:2018jma, Elder:2019yyp,Jenke:2020obe}. 
For chameleons, the mass in dense environments is given by
\be 
m^2(\rho_m)= n(n+1) \frac{\Lambda^{n+4}}{\phi^{n+2}(\rho_m)}~,\ \ \phi(\rho_m)= \left( \frac{n\Lambda^{n+4}m_{\rm Pl}}{\beta \rho_m} \right)^{1/n+1}~.
\ee
For a given $n$, the scale $\Lambda$ and the constant coupling $\beta= m_{\rm Pl}/M$ are constrained by the wide range of experiments mentioned in the introduction. For a recent review see~\cite{CANTATA:2021ktz}.

\section{The anomalous spin precession}
\label{sec:classical-precession}
The anomalous magnetic moment of a particle such as the muon, or the anti-muon, can be determined by comparing the cyclotron frequency of the particle in a constant magnetic field to the spin precession frequency. Scalar interactions modify both frequencies and can therefore contribute to the experimental determination of the anomalous magnetic moment. In effect, this is a classical contribution of the scalar field which must be added to quantum effects arising from radiative corrections akin to the ones obtained from Standard Model particles. We will discuss below how both this classical effect and the quantum contribution can relieve the tension between the Fermilab (and BNL) measurement and the SM prediction on the $(g-2)$ measurements. We first show how the scalar field modifies the cyclotron frequency before considering the effect of the scalar on the spin precession of a particle. We then derive the anomalous spin precession  in detail and show how the effect of the scalar field is included, showing that the contribution of the scalar field is from the Thomas precession via the scalar acceleration. This section is rather technical and some readers may wish to see only the final result in Eqs.~\eqref{anomalous-frequency},~\eqref{frequency-scalar}.  We also present a simplified derivation with only the essential points in Appendix~\ref{appendix:frequency}.

\subsection{Cyclotron frequency}

We are interested in charged relativistic particles moving in an external electromagnetic field and a background scalar field profile. The dynamics of a point particle are governed by the Lagrangian \cite{Uzan:2020aig}
\be
S=\int d\tau( -m A(\phi) \sqrt{-u_E^2} +q A^\mu u^E_\mu)~.
\ee
Notice that the mass of the particle is taken to be effectively scalar-dependent $m_E= m A(\phi)$. This follows from the coupling of the scalar field to matter. The coupling function $A(\phi)$ will be explicitly specified when we discuss the effects of chameleons and symmetrons   on the anomalous spin precession, but is left general for now. 
The associated Euler-Lagrange equation is given by
\be
m_E\frac{du^E_{\mu}}{d\tau}= -q u^\nu_E F_{\nu\mu} -m_E \partial_\phi \ln A \partial_\mu \phi~,
\label{eul}
\ee
where $u^\mu_E=\frac{dx^\mu}{d\tau}$ is the velocity 4-vector of the particle, and $\tau$ is the particle's proper time
$
d\tau= \gamma^{-1} dt
$
where the Lorentz factor is defined as usual by
$
\gamma=\frac{1}{\sqrt{1-v^2}}.
$
On shell we have $u^2_E=-1$. The label $E$ refers to the Einstein frame where space-time is defined by the Minkowski metric and the mass of the particle is field dependent. 
The velocity of the particle is $v^a= \frac{dx^a}{dt}$ where $a$ denotes spatial indices. 
Let us focus on static scalar and magnetic fields ($ F_{ab}= \epsilon_{abc} B_c $), as befitting experiments where the anomalous magnetic moment of muons is measured. We also include an electric field $E^a= F^{0a}$. In the absence of an electric field we have $\frac{du^0_E}{d\tau}=0$ where $u^0_E= \gamma$, implying that $\gamma$ is constant.
With an electric field we have
\be
m_E\frac{du^0_E}{d\tau}= m_E\gamma \frac{d\gamma}{dt}= q\gamma v^a E_a~,
\ee
implying that $\gamma$ is no longer constant and the speed varies unless $v^a E_a=0$, i.e. the trajectory is perpendicular to the electric field. In fact this is precisely the electric field configuration which corresponds to the experimental setting.   Otherwise, we have the non-conservation equation
\be
m_E\frac{d\gamma}{dt}= qv^a E_a~.
\ee
The $(g-2)$ experiments involve a quadrupolar electric field such that $\vec v \cdot \vec E=0$ and therefore $\gamma$ is constant.
The equation of motion is then simply
 in vector notation
\be
\frac{d\vec v}{dt}= -\frac{q}{m_E\gamma} \left( \vec E \cdot \vec v \right) \vec v + \frac{q}{m_E\gamma} \vec E + \frac{q}{m_E\gamma} \vec v\wedge \vec B - \frac{\partial_\phi \ln A}{\gamma^2} \vec\nabla \phi~,
\ee
Using the expansion
$
\vec v\wedge (\vec v \wedge \vec E)= (\vec v \cdot \vec E) \vec v - v^2 \vec E
$
and,  like  in the BNL~\cite{Muong-2:2006rrc} and Fermilab~\cite{Muong-2:2021ojo} experiments, taking the electric field to be perpendicular to the trajectory, i.e.  $\vec v \cdot \vec E=0$,  then the relativistic Newton's law can be written as a gyroscope equation
\be
\frac{d\vec v}{dt}= \vec \omega_c \wedge \vec v~,
\ee
where the angular velocity vector is defined by 
\be
\vec \omega_c= -\frac{q}{m_E\gamma} \vec B + \frac{q}{m_E\gamma v^2} \vec v \wedge \vec E -\frac{\partial_\phi \ln A}{v^2 \gamma^2}\vec v \wedge \vec \nabla \phi~.
\ee
We have assumed that the scalar field gradient is perpendicular to the trajectories $\vec v \cdot \vec \nabla \phi=0$. The last term comes from the scalar interaction with the particle and has a structure similar to the one coming from the Coulomb interaction of a particle with an electric field. 
In the following we will often use the identity
$
\frac{1}{v^2}=\frac{\gamma^2}{\gamma^2-1}
$
implying that 
\be
\vec \omega_c= -\frac{q}{m_E\gamma} \vec B + \frac{q\gamma}{m_E(\gamma^2-1)} \vec v \wedge \vec E -\frac{\partial_\phi \ln A}{\gamma^2-1}\vec v \wedge \vec \nabla \phi~.
\label{cyclotron-frequency}
\ee
Notice that in the ultra-relativistic limit $\gamma\to \infty$, the scalar contribution is suppressed relative to the electric one. In the models that we will consider $A(\phi)\simeq 1$ implying that $m_E\simeq m$. The only effect of the scalar field appears in the extra $\partial_\phi \ln A$ term.

\subsection{Effect of the scalar field on the spin precession of a particle}

We now come to the effect of the scalar field on the spin precession of a relativistic point particle. We will first characterise it 
in the absence of electric and magnetic fields (see \cite{Brax:2020vgg, Brax:2021qqo}). The electromagnetic effects will be discussed in the following section.  The spin vector of the particle evolves according to  
\be
\dot S^{\mu\nu}= 2 p^{[\mu} u^{\nu]}~,
\ee
where $p^\mu = m u^\mu$ at leading order and $\dot\  = u^\mu \nabla_\mu$. 
We work in the Jordan frame here, i.e. the frame where the particle interacts with the gravitational field only, and $\nabla_\mu$ is the covariant derivative associated to the Jordan frame metric $g_{\mu\nu}$. In this frame, gravity is non-canonical and the Einstein-Hilbert term of the gravitational action involves a field-dependent Newtonian gravitational constant. On the other hand,  the Einstein frame metric where Newton's constant is field independent is assumed to be the one of Minkowski space, as we take no gravitational effects into account here. Moreover we will assume that the Jordan and Einstein metrics are conformally related only. 
Here the Pauli-Lubanski vector representing the spin of the particle is given by
\be
S_\mu=-\frac{1}{2} \epsilon_{\mu\nu\rho\sigma} u^\nu S^{\rho\sigma}.
\ee
It is more convenient to study the spin precession in  the local frame which moves with the body. This is achieved in two steps. First by going to the local Minkowski frame (static) in space-time located at the point where the particle is. This is realised by transforming vectors via the vielbein $e^{i}_\mu$ where $i$ is a local Lorentz index, i.e. $f^i=e^i_\mu f^\mu$ for a vector $f^\mu$. Then one needs to apply  a boost $\Lambda^i_{\hat j}$ taking to the hatted frame where the particle is locally at rest, i.e $f^{\hat i}= \Lambda^{\hat i}_i f^i$.
We have the explicit expressions for this boost
\be
\Lambda^0_{\hat 0}= u^0,\ \  \Lambda^0_{\hat a}= u_a, \ \Lambda^a_{\hat b}= \delta^a_b + \left( \gamma-1 \right) \frac{v^av_b}{v^2}.
\ee
We can now define the matrix which transforms vectors in space-time to vectors in the local frame attached to the particle
\be
e^\mu_{\hat i}= \Lambda^i_{\hat i} e^\mu_i~,
\ee
and the spin in the local frame becomes
\be
S_{\hat i}= e^\mu_{\hat i} S_\mu~,
\ee
The spin equation is then
\be
\frac{dS_{\hat i}}{d\tau_J}= S_\mu u^\nu \nabla_\nu e^\mu_{\hat i}~,
\ee
where the covariant derivatives $\nabla_\mu$ are calculated in the Jordan frame with the Jordan metric. { Notice that here $\tau_J$ is the particle's proper time in the Jordan frame.}
Using the orthogonality condition
$
g_{\mu\nu} e^\mu_{\hat i } e^\nu_{\hat j}=\eta_{\hat i \hat j}
$
we find that the covariant derivative involves an antisymmetric tensor
$
u^\nu \nabla_\nu e^\mu_{\hat i}= -\omega_{\hat i} ^{\hat j} e^{\mu}_{\hat j}
$
which is related to the spin precession vector by
$
\omega_{\hat a}= \epsilon_{\hat a\hat b \hat c} \omega^{\hat b\hat c}
$
leading to the spin precession equation 
\be
\frac{d\vec S}{d\tau_J}= \vec \omega\wedge \vec S~,
\ee
confirming that $\omega_{\hat a}$ is the spin precession vector in the local frame moving with the body.

In practice the particle couples to  the Jordan metric
$
g_{\mu\nu}= A^2(\phi) \eta_{\mu\nu}
$
as we neglect the gravitational effects locally. So we have $e^\mu_i= A^{-1}(\phi) \delta^\mu_i$ and therefore
\be
e^\mu_{\hat i}= A^{-1}(\phi) \Lambda^i_{\hat i}\delta^\mu_i.
\ee
The spin precesses according to 
$
\frac{d S_{\hat i}}{d\tau_J}= -\omega_{\hat i \hat j}S^{\hat j}~,
$
where
$
\omega_{\hat i \hat j}= g_{\mu\nu} e^{\mu}_{\hat i} u^\rho \nabla_\rho e^\nu_{\hat j}.
$
We have the explicit expression for the covariant derivative
$
\nabla_\rho e^{\nu}_{ j}= \partial_\rho e^{\nu}_{ j} - \Gamma^\nu_{\rho \mu} e^{\mu}_{ j}
$
implying that
\be
u^\rho \nabla_\rho e^\nu_{\hat j}= e^\nu_{\hat k} \Lambda ^{\hat k}_i \dot \Lambda^i_{\hat j} + u^\rho \left( \partial_\rho e^{\nu}_{ j} - \Gamma^\nu_{\rho \mu} e^{\mu}_{ j} \right) \Lambda^j_{\hat j}~,
\ee
with $\dot \Lambda^i_{\hat j}= u^\rho\partial_\rho \Lambda^i_{\hat j}$.
We have also for the Christoffel symbols
$
\Gamma^{\nu}_{\rho\mu}= \partial_\rho \ln A \delta^\nu_{\mu} + \partial_\mu \ln A \delta^\nu_{\rho}- \partial^\nu \ln A \eta_{\mu\rho}
$~,
and therefore the covariant derivative
$
\nabla_\rho e^{\nu}_{ j}= \partial_\mu \ln A \delta^\mu_{ j} \delta^\nu_\rho - \partial^\nu \ln A \delta_{j \rho}
$
giving finally for  the expression of  the local precession matrix
\be
\omega_{\hat i \hat j}= \partial_\phi A \left( e^{\mu}_{\hat i} u_{\mu} \partial_\nu \phi e^{\nu}_{\hat j}- e^{\mu}_{\hat j } u_{\mu} \partial_\nu \phi e^{\nu}_{\hat i} \right) +
\eta_{\hat i \hat k} \Lambda ^{\hat k}_i \dot \Lambda^i_{\hat j}~.
\label{spinprec}
\ee
The first term comes from the intrinsic coupling between the scalar field and the spin whereas
the last term is due to the change of frame to the local frame where the particle is at rest. We will evaluate it later and relate it to the Thomas precession.

It is also illustrative to boost back the spin vector to the local Minkowski frame where the particle moves with the velocity vector $v^a$ in the Einstein frame.
The two vectors are related by
$
S_{\hat i} = \Lambda^{i}_{\hat i} S_i
$
where $S_i$ is the spin in the laboratory frame, 
we find that for the spatial part
\be
\frac{d\vec S}{d\tau_J}= \vec \omega^\phi\wedge \vec S~,
\ee
$
\omega^\phi_a= \frac{1}{2} \epsilon^{bc}_a \omega^\phi_{bc}
$
and
$
\omega_{ i  j}^\phi= \partial_\phi A ( e^{\mu}_{ i} u_{\mu} \partial_\nu \phi e^{\nu}_{ j}- e^{\mu}_{ j } u_{\mu} \partial_\nu \phi e^{\nu}_{ i})
$.
Here $u^\mu$ is the Jordan frame velocity vector such that $u^\mu= A^{-1} u^\mu_E$ and $u^\mu_E=(\gamma, \gamma v^i)$ is the Einstein velocity as measured in the laboratory frame, assumed to be locally Minkowski and at rest.
When the scalar field is static we have the identity 
\be
\vec \omega^\phi= {\partial_\phi A}\gamma\ \vec v \wedge \vec \nabla \phi~.
\ee
This induces the  precession of the particle in the laboratory frame.

Let us come back to  the spin precession effect $\omega_{\phi\hat i \hat j}$ in the boosted frame where the particle is at rest. In the next section, this will be added to the electromagnetic effects
\be
\omega_{\phi \hat i \hat j}=  \partial_\phi A (  u_{\hat i} \partial_{\hat j}\phi -  u_{\hat j} \partial_{\hat i}\phi ) +\eta_{\hat i \hat k} \Lambda ^{\hat k}_i \dot \Lambda^i_{\hat j}~.
\ee
In the moving frame  the particle is at rest implying that $u_{\hat a}=0$. As a result 
\begin{equation}
\omega_{\phi \hat a \hat b}=\eta_{\hat a \hat c} \Lambda ^{\hat c}_i \dot \Lambda^i_{\hat b}
\end{equation}
implying that the only effect due to the scalar field in the rest frame comes from the boost from the laboratory frame. This will be calculated in (\ref{TS}) and will lead to the Thomas precession. In summary, once the Thomas precession effect has been taken into account, no other contribution to the spin precession will arise from the interaction of the scalar field and the particle { in its rest frame}. In the following section, we will add the effects of the electromagnetic interactions.

\subsection{The Bargmann-Michel-Telegdi equation}

We can now take into account the coupling to an electromagnetic field and its effect on the spin precession. 
In an electromagnetic field the spin evolves according to
\be
\frac{dS^\mu}{d\tau}=  \frac{q}{m_E} \left( \frac{g}{2} F^{\mu\nu} S_\nu + \left( \frac{g}{2}-1 \right) u^\mu_E S_\lambda F^{\lambda\nu}u_\nu^E \right)~,
\ee
where we impose $u_\mu^E S^\mu=0$ as an orthogonality condition. These equations are valid in the Einstein frame where the metric is locally Minkowski, hence the presence of $u^\mu_E$ and not $u^\mu$. 
This equation does not include the effect of the coupling of the scalar field to the spin that we described previously. That will be added at the end of this subsection.

It is more convenient to work in the local frame where the particle is at rest after applying the boost $\Lambda^\mu_{\hat i}= \delta^\mu_{i} \Lambda^{i}_{\hat i}$.
In this frame $u^{\hat i}=(1,\vec 0)$ and therefore the local spin  $S_{\hat i}= \Lambda^\mu _{\hat i} S_\mu$ is such that $S_{\hat 0}=0$. The spin equation becomes then
\be
\frac{d S^{\hat i}}{d\tau}=  \frac{q}{m_E} \left( \frac{g}{2} F^{\hat i\hat j} S_{\hat j} + \left( \frac{g}{2}-1 \right) u^{\hat i} S_{\hat k} F^{\hat k\hat j}u_{\hat j} \right) - \frac{d \Lambda_\mu ^{\hat i}}{d\tau} \Lambda^{\mu}_ {\hat j} S^{\hat j}~,
\ee
where the last term is the change of frame effect we have already encountered.
Specialising this equation to spatial directions only we find
\be
\frac{d S^{\hat a}}{d\tau}=  \frac{qg}{2m_E}   F^{\hat a\hat b} S_{\hat b} - \frac{d \Lambda_\mu ^{\hat a}}{d\tau} \Lambda^{\mu}_ {\hat b} S^{\hat b}~,
\ee
where we  explicitly obtain 
\be
\frac{d \Lambda_\mu ^{\hat a}}{d\tau} \Lambda^{\mu}_ {\hat b}= \frac{\gamma^2}{\gamma +1} \left( \frac{dv_{\hat b}}{d\tau} v^{\hat a} - \frac{dv^{\hat a}}{d\tau} v_{\hat b} \right)~.
\label{TS}
\ee
Combining these terms we can write the spin equation as 
\be
\frac{d S^{\hat a}}{d\tau}= -\omega^a_{s \hat b} S^{\hat b}~,
\ee
with the angular velocity matrix
\be
\omega^{\hat a}_{s \hat b}= -\frac{qg}{2m_E}  F^{\hat a}_{\hat b}+ \frac{\gamma^2}{\gamma +1} \left( \frac{dv_{\hat b}}{d\tau} v^{\hat a} - \frac{dv^{\hat a}}{d\tau} v_{\hat b} \right)~.
\ee
In the boosted frame  the spatial parts of the electromagnetic field strength becomes
$
 F^{\hat a\hat b}= \epsilon^{\hat a \hat b \hat c} B_{\hat c}
$
where the boosted magnetic field is given by 
\be
B_{\hat a}= \left[\gamma \left( B_a - (\vec v \wedge \vec E)_a \right) - \frac{\gamma-1}{v^2} (\vec B \cdot \vec v) v_a \right]\delta^{a}_{\hat a}~,
\ee
which leaves invariant the component parallel to the velocity. 
Defining the spin precession vector as 
$
\omega_{s}^{\hat a}= \frac{1}{2\gamma} \epsilon^{\hat a \hat b\hat c} \omega_{s\hat b\hat c}
$
we find that this reduces to 
\be
\omega_{s}^{\hat a}= -\frac{qg}{2m_E\gamma}B^{\hat a}+  \frac{\gamma^2}{\gamma+1} \left( \vec a \wedge \vec v \right)^a \delta^{\hat a}_a~.
\label{spins}
\ee
This is the precession vector for a spinning particle subject to an electromagnetic field. The last term comes from the change of frame effect
in Minkowski space and is nothing but the  Thomas precession vector
\be
\omega_{T}^{\hat a}=  \frac{\gamma^2}{\gamma+1} (\vec a \wedge \vec v)^a \delta^{\hat a}_a~,
\ee
involving the acceleration of the particle
\be
\vec a= \frac{d\vec v}{dt}~,
\ee
containing both the scalar and electromagnetic forces. Notice that this is to be evaluated in the laboratory frame.
The spin equation reads now
\be
\frac{d \vec S}{dt}= \vec\omega_s \wedge \vec S~,
\ee
in the  frame moving  with the particle. This includes all the electromagnetic and scalar effects on the spin precession of a particle as the intrinsic contribution discussed in the previous section vanishes identically.

\subsection{Comparing the spin precession  to the angular velocity}

\subsubsection{Boosting the angular velocity} 

We are interested in the anomalous spin precession defined as the difference between the spin precession and the angular velocity vector in the moving frame. 
We will compare the spin precession in the boosted frame as obtained in (\ref{spins}) to the angular velocity vector in the boosted frame too.

Let us now boost the angular velocity vector to the moving frame and compare it to the spin precession vector. To do this, recall that the vector
\be
\vec \omega_c= -\frac{q}{m_E\gamma} \vec B + \frac{q\gamma}{m_E(\gamma^2-1)} \vec v \wedge \vec E -\frac{\partial_\phi \ln A}{\gamma^2-1}\vec v \wedge \vec \nabla \phi~,
\ee
contains two parts
\be
\vec \omega_c= \vec \omega_B + \vec \omega_E~.
\ee
The first part is a vector proportional to the magnetic field
\be
\vec \omega_B= -\frac{q}{m_E\gamma} \vec B~.
\ee
Covariantly (see Appendix \ref{app:tenso})   $\vec \omega_B$ is the magnetic part of a two-index antisymmetric tensor $\omega_{B i  j}=-\frac{q}{m} B_{ i j}$ representing the part of the angular rotation of the particle in the laboratory frame
$\omega_B^{ a}= \frac{1}{2} \epsilon^{ a  b  c} \omega_{B  a  b}$. As a result this transforms under a Lorentz boost as the magnetic part of $\omega_{B i  j}$
\be
 \omega^{\hat a}_B= \left (\gamma \omega^{ a}_B - \frac{\gamma^2}{\gamma+1} \left(\vec v \cdot \vec \omega_B \right) v^a \right)\delta^a_{\hat a}~,
\ee
under a Lorentz boost to the moving frame. 
The second part is a vector 
\be 
\vec \omega_E=\frac{q\gamma}{m_E(\gamma^2-1)} \vec v \wedge \vec E -\frac{\partial_\phi \ln A}{\gamma^2-1}\vec v \wedge \vec \nabla \phi~,
\ee
orthogonal to $\vec v$. Covariantly this is associated to the antisymmetric tensor $E_{ij}=\epsilon_{ijk} {\cal E}_k$ determined by the spatial vector ${\cal E}_a$. Applying a boost on $E_{ij}$ as $E_{\hat i \hat j}= \Lambda_{\hat i}^i\Lambda_{\hat j}^j E_{ij}$, we find that the  vector ${\cal E}_a$ is invariant implying that $ \omega_{E\hat a }\equiv \omega_{Ea}\delta^a_{\hat a}$, i.e the ``electric'' part of the angular velocity is invariant under a Lorentz boost to the moving frame. 

Collecting these two terms we find  more explicitly
\be
\omega_c^{\hat a}= \left[ -\frac{q}{m_E} \left(B^a -\frac{\gamma}{\gamma+1} \left(\vec v \cdot \vec B \right) v^a \right) + \frac{q\gamma}{m_E(\gamma^2-1)} \vec v \wedge \vec E -\frac{ \partial_\phi \ln A}{\gamma^2-1}\vec v \wedge \vec \nabla \phi \right]\delta^{\hat a}_a~,
\ee
where the boost of the magnetic part is explicit. 

\subsubsection{The Thomas precession and the spin precession}

The spin precession vector in the moving frame depends crucially on the Thomas precession. This part of the precession is obtained upon  using the acceleration in the laboratory frame
\be
\vec a =   \frac{q}{m_E\gamma} \vec E + \frac{q}{m_E\gamma} \vec v\wedge \vec B - \frac{\partial_\phi \ln A}{\gamma^2} \vec\nabla \phi~,
\ee
leading to the Thomas precession
\be
\omega_{T}^{\hat a}=  \left[- \frac{q\gamma}{m_E(\gamma+1)} \vec v \wedge \vec E + \frac{\partial_\phi \ln A}{\gamma +1} \vec v \wedge \vec\nabla \phi \right]^a \delta^{\hat a}_a~.
\ee
Notice that there is a term coming from the scalar field on par with the usual electric field contribution. All in all, we get for the spin precession vector in the moving frame
\begin{align} \nonumber
\omega_{s}^{\hat a} = -\frac{q}{m_E} \bigg[ \frac{g}{2} &\left( B^{\hat a} - \frac{\gamma}{\gamma+1} (\vec B \cdot \vec v) v^{\hat a} \right) - \left( \left(\frac{g}{2}-1 \right) + \frac{1}{\gamma+1} \right) (\vec v \wedge \vec E)^{\hat a}  \bigg] \\
&+ \frac{\partial_\phi \ln A}{\gamma +1} \vec v \wedge \vec\nabla \phi^{\hat a}~,
\end{align}
where the influence of the scalar field only appears from the Thomas precession term  via the acceleration equation.

\subsubsection{The anomalous spin precession}

Putting it all together we obtain for the difference between the spin precession and the angular velocity vectors in the moving frame
\begin{align} \nonumber
\vec \omega_s -\vec \omega_c = -\frac{q}{m_E} &\left[ a_\mu \left(\vec B - \frac{\gamma}{\gamma+1} (\vec B \cdot \vec v) \vec v \right) + \left(a_\mu -\frac{1}{\gamma^2-1} \right) \vec v \wedge \vec E \right] \\
&+\frac{\gamma}{\gamma^2-1}\partial_\phi \ln A \vec v \wedge \vec \nabla \phi~,
\label{BMT}
\end{align}
where $a_\mu= \frac{g}{2}-1$\footnote{ Notice that this notation is traditional although rather unfortunate as the index $\mu$ refers to the muons and not a Lorentz index.}.
We retrieve the anomalous spin precession equation when no scalar is present 
\be
\vec \omega_a \equiv \vec \omega_s -\vec \omega_c = -\frac{q}{m_E} \left[ a_\mu  \left(\vec B - \frac{\gamma}{\gamma+1} (\vec B \cdot \vec v) \vec v \right) + \left(a_\mu -\frac{1}{\gamma^2-1} \right) \vec v \wedge \vec E \right]~.
\label{anomalous-frequency}
\ee
The scalar field corrects the anomalous spin precession  by an amount
\be
\delta \vec \omega_a= \frac{\gamma}{\gamma^2-1}\partial_\phi \ln A \vec v \wedge \vec \nabla \phi~,
\label{frequency-scalar}
\ee
which combines the effects of the Thomas precession and the scalar force acting on the particle. {A simpler derivation in the absence of any electric field is given in Appendix \ref{appendix:frequency}.}

\section{Quantum Corrections}

In this section we compute the quantum corrections to the anomalous magnetic moment arising from the coupling of the scalar field to fermions. This was computed in detail within the context of electron magnetic moment experiments \cite{Brax:2018zfb}, so we only recount the main points here. We first show the coupling of the scalar field to fermions and photons before computing the loop corrections. 

\subsection{Couplings to fermions and photons}
\label{sec:L}
 From the Lagrangian in equation (1) we see that quantum contributions to the anomalous magnetic moment arise from the direct coupling between the scalar field and fermions.  
 Massive fermions, such as the muon, obey the modified Dirac equation \cite{Brax:2010jk}
\begin{equation}
\label{eq:Dirac}
\mathcal L_m \supset \bar\psi[i\slashed{D} - A(\phi) m_\mu]\psi~,
\end{equation}
where $D_\mu = \partial_\mu + i e A_\mu$ is the gauge-covariant derivative. Note that $A(\phi)$ is the coupling between the scalar and muon fields, while $A_\mu$ is the photon field.   The mass of the fermion now depends on the scalar field. In addition the scalar could also couple to photons as this is not explicitly forbidden by the symmetries of the theory \cite{Brax:2007hi, Burrage:2008ii, Brax:2009ey,Brax:2010uq}. The coupling function could be different to that of the fermion $\varepsilon(\phi)$, which modifies the kinetic term of the photon to read
\begin{equation}
\label{eq:defVarepsilon}
\mathcal L_m \supset - \frac{1}{4}(1+\varepsilon(\phi)) F_{\mu\nu} F^{\mu\nu}.
\end{equation}
As both $A(\phi)$ and $\varepsilon(\phi)$ introduce non-renormalizable operators in the models, these theories should be viewed as low-energy effective field theories (EFTs) valid only below some cutoff. These models typically satisfy $A(\phi) \approx 1$ and $\varepsilon(\phi) \approx 0$.  At the linearized level their effects are captured by the  dimensionless coupling strengths
\begin{equation}
\beta_m(\phi) = m_{\rm Pl}\frac{\text{d}\ln A}{\text{d}\phi}~,
\quad
\beta_\gamma(\phi) = m_{\rm Pl}\frac{\text{d}\ln(1+\varepsilon)}{\text{d}\phi}~\simeq m_{\rm Pl}\frac{\text{d}\varepsilon}{\text{d}\phi}
\end{equation}
where the last equality follows if we assume that $\varepsilon\ll 1$.
These couplings are typically at least as strong as gravity, or $\beta_m, \beta_\gamma \geq 1$.
In the following, we will focus on the matter coupling.  For the chameleon, bounds from particle physics impose that $M\gtrsim 10^2$ GeV \cite{Brax:2009aw}.

\subsection{Loop corrections}

Let us consider quantum fluctuations $\chi = \phi - \bar\phi$ about the classical background field profile $\bar\phi$ in the experiment where $g-2$ is to be measured. As the muons remain very close to the center of the experimental cavity, it suffices to take $\bar\phi \approx \phi_0$ to be a constant, where $\phi_0$ is the classical field value at the center.
 At the one-loop level, the only influence from $V(\phi)$ is a mass term for the $\chi$ field, with mass $m_0$ given by the second derivative
\begin{equation}
\label{eq:def_mphi}
m_0^2 = V_{\text{eff},\phi\phi}(\phi_0)~,
\end{equation}
evaluated at the center of the cavity. Note that we can linearise Eqs.~\eqref{eq:Dirac} and \eqref{eq:defVarepsilon} and we obtain the leading order terms~\cite{Brax:2009ey,Brax:2009aw}
\begin{equation}
\label{eq:L_linear_couplings}
\mathcal L_m \supset - \left( \frac{\beta_m m_\mu}{m_{\rm Pl}} \right) \bar\psi\psi \chi - \frac{1}{4}\left(\frac{\beta_\gamma}{m_{\rm Pl}}\right) \chi F_{\mu\nu} F^{\mu\nu},
\end{equation}
where $\beta_m \equiv \beta_m(\phi_0)$ and $\beta_\gamma \equiv \beta_\gamma(\phi_0)$. 

\begin{figure}
\includegraphics[width=70mm]{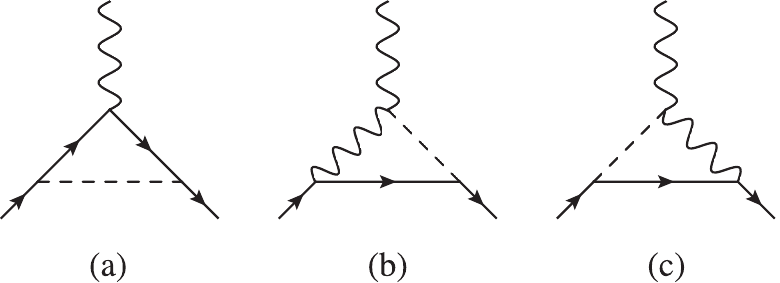}
\caption{\scriptsize Scalar field (dashed line) contributions at one-loop order to the magnetic moment of the muon.  Reproduced from \cite{Brax:2018zfb}.}
\label{fig:FeynmanDiagrams}
\end{figure}

At one-loop order there are three Feynman diagrams to consider for the anomalous magnetic moment, as in Fig.~\ref{fig:FeynmanDiagrams}. Here we quote the results since these diagrams have appeared widely in the literature (see, e.g., Refs.~\cite{Giudice:2012ms,Jegerlehner:2009ry,PhysRevD.5.2396,Chen:2015vqy,Marciano:2016yhf}, as well as \cite{Arcadi:2021zdk} for recent applications to models of dark matter). 
The first diagram in Fig.~\ref{fig:FeynmanDiagrams}(a) gives the finite contribution
\begin{equation}
\delta a \supset 2\beta_m^2 \left(\frac{m_\mu}{4\pi m_{\rm Pl}}\right)^2 I_1(m_0/m_\mu)~,
\label{amu-correction-general-qm}
\end{equation}
whereas the remaining two diagrams are UV divergent and require renormalisation. Using the $\overline{\text{MS}}$ scheme gives
\begin{equation}
\label{eq:da_bMbG}
\delta a \supset 4\beta_m\beta_\gamma\left(\frac{m_\mu}{4\pi m_{\rm Pl}}\right)^2
\left[ \log\left(\frac{\mu}{m_\mu}\right) + I_2(m_0/m_\mu)\right],
\end{equation}
where $\mu$ is an arbitrary energy scale. The two finite integrals above are
\begin{subequations}
\label{eq:feyn_integrals}
\begin{align}
\label{eq:feyn_integral_1}
I_1(\eta) &= \int_0^1\text{d}x \frac{(1-x)^2(1+x)}{(1-x)^2+x \eta^2}~,
\\
\label{eq:feyn_integral_2}
I_2(\eta) &= \int_0^1\text{d}x\int_0^1\text{d}y (x-1)\log[x^2 + (1-x)y \eta^2]~;
\end{align}
\end{subequations}
where $\eta=\frac{m_{0}}{m_{\mu}}$.
In the regime of interest we typically have $m_0 \ll m_\mu$, such that it suffices to set $m_0/m_\mu = 0$ in the integrals and we have
\begin{equation*}
I_1(0) = I_2(0) = \frac{3}{2}~.
\end{equation*}
In Eq.~\eqref{eq:da_bMbG} there is an arbitrary scale $\mu$. In low-energy physics there is an ambiguity in determining this scale and a conservative estimate is to set
$
\log(\mu/m_\mu) \sim 1
$
in order to evaluate the order of magnitude of the correction term. Here, though, we consider the dominant contribution and set $\beta_\gamma=0$, giving the quantum correction to be
\begin{equation}
    \delta a= 3\beta_m^2 \left(\frac{m_\mu}{4\pi m_{\rm Pl}}\right)^2 I_1 \left( \frac{m_0}{m_\mu} \right),
    \label{quantum-contribution}
\end{equation}
which should be added to the classical contributions already discussed.

\section{The anomalous magnetic moment of the muon}
\subsection{A story of signs }

\begin{figure}[t]
    \centering
    \includegraphics[width=0.3\textwidth]{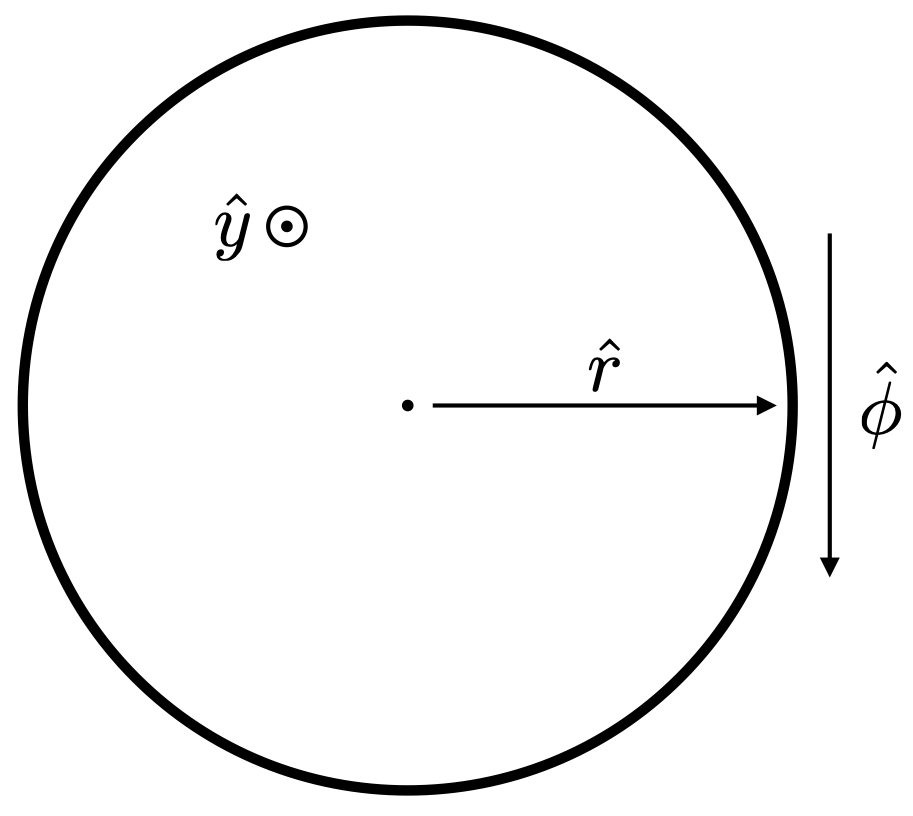}
    \caption{\scriptsize Definition of the coordinate system on the storage ring, matching the conventions of \cite{Albahri:2021kmg}.}
    \label{fig:coordinates}
\end{figure}

In this section we will compare the relative signs of the terms in the anomalous spin precession  to show that a new scalar field can alleviate, and not exacerbate, the tension between experiment and theory.  To start, we first recall that the anomalous spin precession is
\begin{equation}
    \vec \omega_a = - \frac{q \vec B}{m_\mu} a_\mu + \frac{\gamma}{\gamma^2 - 1} \partial_\phi \ln A \vec v \wedge \vec \nabla \phi~,
\end{equation}
where we have dropped the two terms that are approximately zero for this particular experiment. Here $m_\mu$ is the mass of the muon in the Einstein frame. Let us ignore the second term on the right hand side for now.  The quantities $q, a_\mu,$ and $m_\mu$ are all positive.  If we assume that the muons are travelling in the clockwise direction (see Fig.~\ref{fig:coordinates}) then the magnetic field points in the $+ \hat y$ direction. 

Next we examine the scalar term.  The field value is large in the center and small near the walls. This follows from the fact that the scalar field takes smaller values in the presence of dense matter for the screened models that we consider, i.e. chameleons and symmetrons.  Furthermore, the antimuons are displaced in the $+ \hat r$ direction from the center.  Thus, the field gradient, at the location of the muons, points in the $- \hat r$ direction.  This implies that the entire second term points in the $- \hat y$ direction.

Since both terms point in the $- \hat y$ direction, we can examine only the $\hat y$ component of the equation:
\begin{equation}
    \omega_a = \frac{q B}{m_\mu} a_\mu + \frac{\gamma}{\gamma^2 - 1} \partial_\phi\ln  A v  |\vec \nabla \phi|~.
\end{equation}
The experimental value may be isolated as
\begin{equation}
    a_{\mu,\mathrm{exp}} = \frac{m_\mu}{q B} \left(\omega_a - \frac{\gamma}{\gamma^2 - 1} \partial_\phi \ln A v  |\vec \nabla \phi| \right)~.
\end{equation}
Without the scalar field correction term in this equation, one would infer a value of $a_\mu$ that is too large as the scalar field contribution is positive. 

At the same time, the theoretical value of the anomalous magnetic moment is a combination of the Standard Model contribution and the quantum mechanical contribution from the new scalar:
\begin{equation}
    a_{\mu,\mathrm{th}} = a_{\mu,\mathrm{SM}} + a_{\mu, \phi\mathrm{QM}}~,
\end{equation}
where $a_{\mu, \phi\mathrm{QM}}$ is given by Eq.~\eqref{amu-correction-general-qm}~. 
Without the scalar correction to the theoretical prediction one would calculate a theoretical value that is too small. 

The experimental value of $a_\mu$ is indeed larger than the theoretical prediction, by an amount
\begin{equation}
    \delta a_\mu = a_{\mu,\mathrm{exp}}-a_{\mu,\mathrm{th}} = 2.51 \times 10^{-9}
\end{equation}
A new scalar field can alleviate the tension if
\begin{equation}
    \delta a_\mu \approx \frac{m_\mu}{q B} \frac{1}{\gamma v} \partial_\phi \ln A  |\vec \nabla \phi| + 2 (\partial_\phi \ln A)^2 \left(\frac{m_\mu}{4 \pi} \right)^2 I_1 \left( \frac{m_0}{m_\mu} \right)~,
    \label{amu-discrepancy}
\end{equation}
where the first and second terms are the classical and quantum corrections to the anomalous spin precession, respectively.  We will explicitly evaluate the magnitude of these terms for a Yukawa theory, chameleons, and symmetrons.

\subsection{Yukawa interaction}

As the simplest example of a scalar particle's interaction with matter, let us first focus on the Yukawa interaction corresponding to a contribution to the particle Lagrangian with a constant coupling strength
\be 
{\delta}{\cal L}= -\frac{\beta}{m_{\rm Pl}} \phi m_\psi \bar \psi \psi~,
\ee
where $\beta$ is this coupling. This corresponds to a modification of gravity with an increase of Newton's constant
\be 
G_N \to (1+2\beta^2) G_N~,
\ee
inside the Compton wavelength of the scalar field, which is taken to be light enough that its range exceeds the size of the solar system. 
The coupling strength is constrained by the Cassini experiment at the $\beta^2 \lesssim 10^{-5}$ level. This could be relaxed as the gluons contribute mainly to the mass of atoms and therefore gravitational tests are mostly sensitive to the coupling of scalars to gluons. Here we illustrate our analysis by taking the coupling strength to be universal to all Standard Model fields. 
The Yukawa interaction is associated with a coupling function
\be 
A(\phi)= 1+ \frac{\beta}{m_{\rm Pl}} \phi~.
\ee
The quantum correction to $(g-2)$ is in this case
\be 
\delta a_\mu= 3 \beta^2 \left(\frac{m_\mu}{4\pi m_{\rm Pl}}\right)^2~.
\ee
Similarly for a laboratory experiment performed on earth, the scalar field profile is mainly due to the earth with a gradient 
\be 
\vec \nabla  \phi= 2\beta m_{\rm Pl} g \vec e_r~,
\ee
where $\vec e_r$ is the radial unit vector from the center of the earth and $g$ the gravitational acceleration in the laboratory. 
The correction to the anomalous spin precession  becomes then 
\be 
\delta \vec \omega_a= \frac{2 \beta^2 g \gamma}{\gamma^2-1} \vec v \wedge \vec e_r~.
\ee
When the antimuons go clock-wise in the experiment which is in the horizontal plane, we find that this contribution goes inwards in the plane towards the center of the ring does. This does not affect the contribution perpendicular to the ring as the one associated to the magnetic field. 
As a result, and as the norm of $\vec \omega_a$ is measured experimentally, this gives a new contribution to the experimental determination of $\omega_a$. Now we find that 
\be 
\frac{\delta \omega_a}{\omega_a}\lesssim 10^{-20}
\ee
after applying the Cassini bound~\cite{Bertotti:2003rm}. Similarly  we find that the quantum contribution from a Yukawa interaction is bounded to be below 
\be 
\delta a_\mu \lesssim 10^{-43}~.
\ee
This is completely negligible. In the following, we will consider screening mechanisms where the Cassini test of deviations from Newtonian law is easily passed whilst the strength of the coupling to matter can be much larger in the context of  laboratory experiments.

\subsection{Chameleon models}

\begin{table}[]
    \centering
    \begin{tabular}{c|c | l}
        $\delta a_\mu$ & $2.51 \times 10^{-9}$ & Discrepancy between experiment and theory\\
        $v$ & 0.999419 & Muons' velocity \\
        $\gamma$ & 29.3  & Muons' Lorentz factor\\
        $B$ & $1.45~\mathrm{T}$ & Magnetic field \\
        $\rho_\mathrm{vac}$ & $10^{-9}~\mathrm{kg~m}^{-3}$ &  Density inside storage ring vacuum\\
        $\rho_\mathrm{wall}$ & $10^4~\mathrm{kg~m}^{-3}$ & Density of storage ring walls \\
        $R_\mathrm{vac}$ & $4.5~\mathrm{cm}$ & Interior radius of storage ring tube
    \end{tabular}
    \caption{\scriptsize Experimental values of Fermilab muon experiment~\cite{Muong-2:2021ojo}.}
    \label{tab:experimental-values}
\end{table}

We have seen that the Yukawa model is too tightly constrained by local tests of gravity to be able to produce the discrepancy between experiment and theory in the muon's anomalous magnetic moment.  A model with screening, however, allows for much larger matter couplings to be compatible with observation.  We begin by considering a canonical example of a screened theory, the chameleon, as given by Eq.~\eqref{chameleon-potentials}.
We first estimate the size of the scalar contribution to Eq.~\eqref{BMT}. 
To proceed we must obtain the gradient of the scalar field $\vec \nabla \phi$.  To this end, we note that inside the walls of the storage ring the field satisfies $\phi \approx 0$. This is the case as the density far exceeds the energy scales associated to the self-interaction potential for chameleons.  Assuming the density of the residual gas inside the storage ring is negligible, the field value rises towards a point where the Compton wavelength is of order the size of the cavity \cite{Hamilton:2015zga}
\begin{equation}
    m_\mathrm{eff}(\phi_\mathrm{vac})^{-1} \approx R_\mathrm{vac}~,
\end{equation}
giving a central field value 
\begin{equation}
    \phi_\mathrm{vac} = \xi \left( n (n + 1) \Lambda^{4+n} R_\mathrm{vac}^2 \right)^{\frac{1}{n+2}}~,
\end{equation}
where $R = 4.5~\mathrm{cm}$ \cite{Danby:2001eh} is the radius of the cavity inside the storage ring, and $\xi$ is a geometry-dependent factor.  Approximating the storage ring as an infinite cylinder, we have $\xi = 0.68$ \cite{Hamilton:2015zga}.  Although the gradient $\vec \nabla \phi$ is not uniform, being steeper towards the edges of the storage ring and shallower towards the center, we will approximate it as simply $|\vec \nabla \phi| \approx \phi_\mathrm{vac} / R$.  It is oriented in the radial direction, perpendicular to the motion of the muons so $\vert \vec v \wedge \vec \nabla \phi\vert  = |\vec v| |\vec \nabla \phi|$~.

Using Eq.~\eqref{amu-discrepancy}, we have
\begin{equation}
    \delta a_\mu = \frac{m_\mu}{q B} \frac{1}{\gamma v} \frac{\phi_\mathrm{vac}}{M R_\mathrm{vac}} + 3 \left( \frac{m_\mu}{4 \pi M}\right)^2~.
    \label{cham-discrepancy}
\end{equation}
We have simplified the quantum term by noting that the chameleons' effective mass $m_\mathrm{eff} = R^{-1}_\mathrm{vac} \approx \mu~\mathrm{eV}$ is very small compared to the mass of the muon,  so we have the limiting case $I_1\left( m_\mathrm{eff} / m_\mu \right) \approx 3/2$~.

The first term is the classical contribution, the second term is the quantum one.  We can only trust the classical term when (i) the perfect vacuum and (ii) the zero skin depth approximations hold, i.e. the approximate vanishing of the scalar field inside the walls.  The perfect vacuum approximation is
\begin{equation}
    \frac{\rho_\mathrm{cav}}{M} \ll \frac{n \Lambda^{4+n}}{\phi_\mathrm{vac}^{n+1}}~,
\end{equation}
and the zero skin depth approximation is
\begin{equation}
    m_0^2 \ll m_\mathrm{wall}^2~,
\end{equation}
where $m^2_\mathrm{eff}(\phi) \equiv V_{\mathrm{eff}, \phi\phi}(\phi)$ is the effective mass of the chameleon, and we have defined $m_0 = m_\mathrm{eff}(\phi_\mathrm{vac})$ and $m_\mathrm{wall} = m_\mathrm{eff}(\phi_\infty)$.  The field value $\phi_\infty$ is the equilibrium deep inside the walls of the storage ring
\begin{equation}
    \phi_\infty = \left( \frac{n \Lambda^{4+n} M}{\rho_\mathrm{wall}} \right)^{1/(1+n)}~.
\end{equation}
The densities are approximately that of a room temperature vacuum at $~1~\mu\mathrm{Torr}$ and steel, respectively~\cite{casey-email} and are given, along with other experimental values, in Table~\ref{tab:experimental-values}.

\begin{figure}[t]
    \centering
    \includegraphics[width=0.5\textwidth]{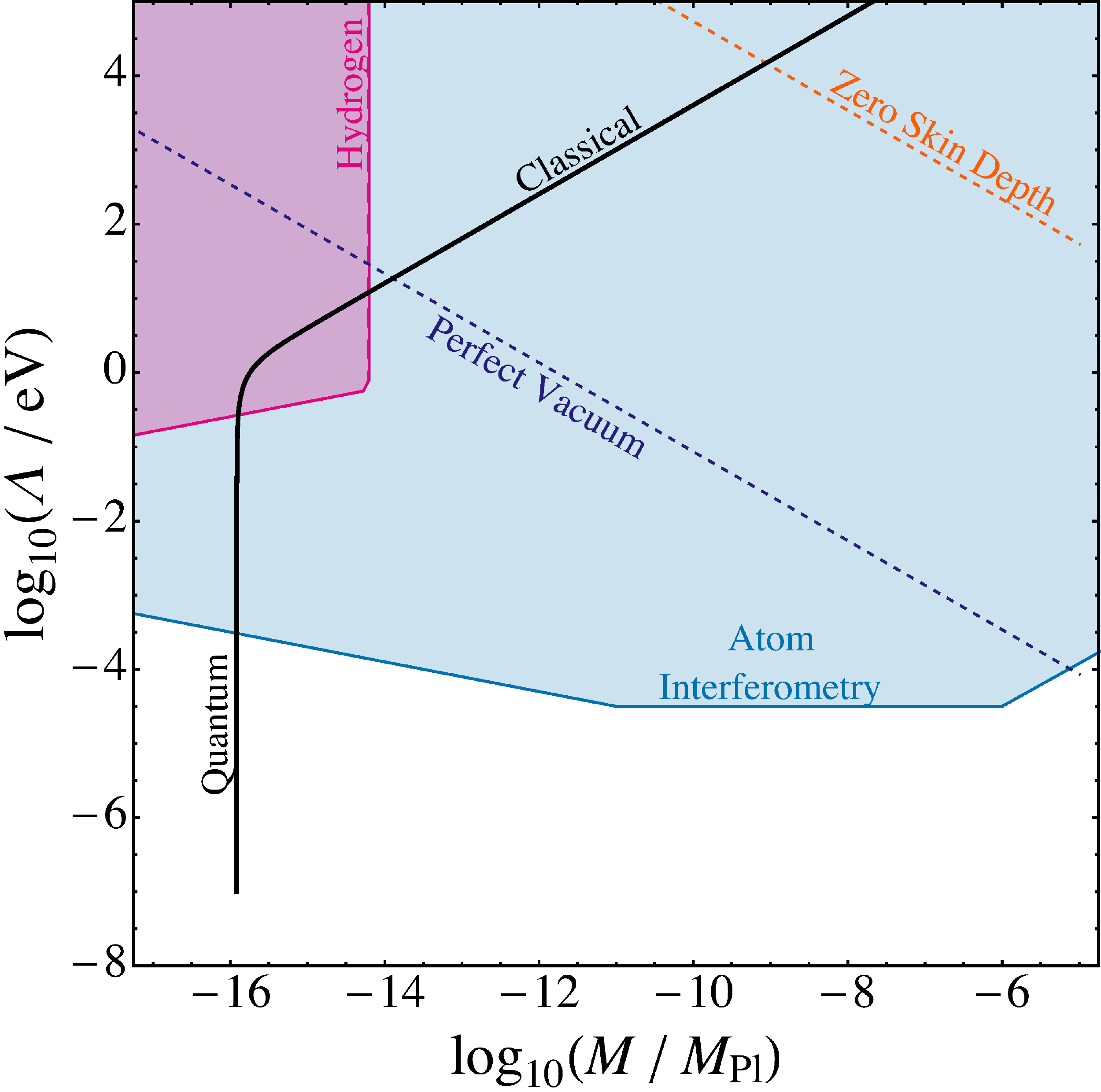}
    \caption{\scriptsize (a) The values of $M, \Lambda$ for the $n = 1$ chameleon that produce the discrepancy between experimental and theoretical values of the muon magnetic moment are plotted with a black line.  The zero skin depth approximation fails in the upper right corner, while the perfect vacuum approximation fails in the lower left half. The areas where the classical and quantum components of Eq.~\eqref{cham-discrepancy} each dominate are indicated.  Only the classical component relies on the perfect vacuum approximation. The hydrogen bounds have been reinterpreted from~\cite{Burrage:2016bwy} to include the screening of the hydrogen nucleus~\cite{Brax:2022olf}.  The region where $M \approx 10^{-16} ~\Mpl,~ \Lambda \lesssim \mathrm{meV}$ remains viable.  The curve stops at $\Lambda \approx 10^{-7}~\mathrm{eV}$ due to the large chameleon mass inside the storage ring. }
    \label{fig:chameleon}
\end{figure}

{
With these assumptions, the values of $\Lambda, M$ that produce the discrepancy $\delta a \equiv |a_\mathrm{exp} - a_\mathrm{SM}| \approx 10^{-9}$ are plotted, along with existing constraints on chameleon parameter space, in Fig.~\ref{fig:chameleon}~. Although the region where the classical effect dominates the quantum one is already already excluded by atom interferometry, there exists a set of parameter values where $M \approx 10^{-16}\Mpl$ and  $\Lambda \lesssim \mathrm{meV}$ for the $n = 1$ chameleon, which could explain the discrepancy $\delta a$.  {The curve terminates at $\Lambda = 10^{-7}~\mathrm{eV}$, as below this point the chameleon mass becomes larger than that of the muon.  This results in $|I_1(m_0 / m_\mu)| \ll 1$, consequently the chameleon's contribution to the anomalous magnetic moment is strongly suppressed.

The viable chameleon parameters are very nearly excluded by collider tests of chameleons which rule out $M \lesssim 10^{-16}~\Mpl$ at $2\sigma$~\cite{Brax:2009aw}.  It is therefore entirely possible that revisiting chameleon constraints with current LHC data would be able to detect or exclude these models.  We leave a detailed analysis of those constraints to future work.}

Note that the bounds deriving from hydrogen spectroscopy~\cite{Brax:2010gp,Burrage:2016bwy} have been reinterpreted and relaxed considerably.  This is because those previous studies did not account for the screening of the proton nucleus or the gas inside the vacuum chamber.  Both effects must be included for the smaller values of $M$ and $\Lambda$ that are of interest here.  The modified hydrogen bounds appear in Fig.~\ref{fig:chameleon}, the details of which are in~\cite{Brax:2022olf}.
}

\subsection{Symmetrons}

For the symmetron model we now perform a similar analysis to the one for chameleons. First we  need to estimate the field $\phi$ and its gradient $\vec \nabla \phi$ at the location of the muons in the storage ring.  To begin, it is helpful to write the effective potential as
\begin{equation}
    V_\mathrm{eff}(\phi) = \frac{1}{2} \left( \frac{\rho}{M^2} - \mu^2 \right) \phi^2 + \frac{\lambda}{4} \phi^4~.
\end{equation}
We will  assume that the density is large in the walls of the storage ring and negligible in the interior:
\begin{equation}
    \rho_\mathrm{vac} \ll \mu^2 M^2 \ll \rho_\mathrm{walls}~.
    \label{crit-density}
\end{equation}
The symmetron field evolves from a value $\phi = 0$ at the surface of the storage ring walls towards its vacuum expectation value (vev) $v \equiv \frac{\mu}{\sqrt \lambda}$.  It naturally does this over approximately one Compton wavelength $\mu^{-1}$, so the relative sizes of the symmetron mass $\mu$ and the radius of the approximately-cylindrical cavity of the storage ring $R_\mathrm{vac} = 4.5$~cm have important consequences for the symmetron phenomenology.  The three possible cases to consider are:
\begin{itemize}
    \item $\mu^{-1} \ll R_\mathrm{vac}$:  The field quickly rolls to the vev a short distance from the vacuum chamber walls and is approximately constant $\phi \approx v$ in the interior of the storage ring.  At the location of the muons we have $\phi \approx v$, $\vec \nabla \phi \approx 0$, and the classical correction to the anomalous spin precession equation is small.  The quantum contribution could still be large, however, and will be considered at the end of this section.
    \item $\mu^{-1} \gg R_\mathrm{vac}$: In this case the field does not have sufficient room to reach the vev, and it is energetically favorable for the field to remain at $\phi = 0$ everywhere inside the storage ring~\cite{Upadhye:2012rc}.  As a result, the classical and quantum corrections to the anomalous spin precession are zero.
    \item $\mu^{-1} \approx R_\mathrm{vac}$: In this case, the field rolls just quickly enough to reach the vev at the center.  As a crude approximation, we have $\phi \approx v$ and $\nabla \phi \approx \frac{v}{R_\mathrm{vac}}$, and the classical correction to the anomalous spin precession  can be significant.
\end{itemize}
For the moment we will focus on this third scenario.  This sets the mass scale of interest to be
\begin{equation}
    \mu \approx R_\mathrm{vac}^{-1} = 4.44 \times 10^{-6}~\mathrm{eV}~.
\end{equation}
In reality, the field does not roll all the way to the vev when $\mu^{-1} \approx R_\mathrm{vac}$, nor is the field gradient linear.  To improve upon the present analysis, one could either (i) perform detailed numerical simulations as was done in for other laboratory experiments~\cite{Elder:2016yxm,Jaffe:2016fsh,Brax:2018zfb,Sabulsky:2018jma,Elder:2019yyp}, or (ii) estimate the central field value as that which sets $m_\mathrm{eff}(\phi_\mathrm{center})^{-1} \approx R_\mathrm{vac}$, as was done with the chameleon, or (iii) rely on the analytic approximation method of \cite{Brax:2018zfb}.  The present analysis represents a best-case scenario, for the classical symmetron correction to the anomalous precession to be as large as possible.  We will find that such an interaction is already constrained by existing experiments, so the simplified analysis presented here is sufficient. 

We will assume  that $\phi = v$ and $\nabla \phi = v / R_\mathrm{vac}$ at the location of the muons.  In natural units, then, the classical contribution to the muons' anomalous magnetic moment determination is
\begin{align} \nonumber
    \delta a_\mathrm{cl} 
    &= \frac{m}{q B} \frac{|\vec v|}{R_\mathrm{vac}} \frac{\mu^2 }{M^2 \lambda}~\sim \frac{m}{q B} \frac{|\vec v|}{R_\mathrm{vac}^3} \frac{1}{M^2 \lambda}~, \\
    &= \left( 1.07 \times 10^{-10} ~\mathrm{eV}^2 \right) \frac{1}{M^2 \lambda}~,
\end{align}
where we have used $\mu \approx R_\mathrm{vac}^{-1}$.

\begin{figure}[t]
    \centering
    \includegraphics[width=0.5\textwidth]{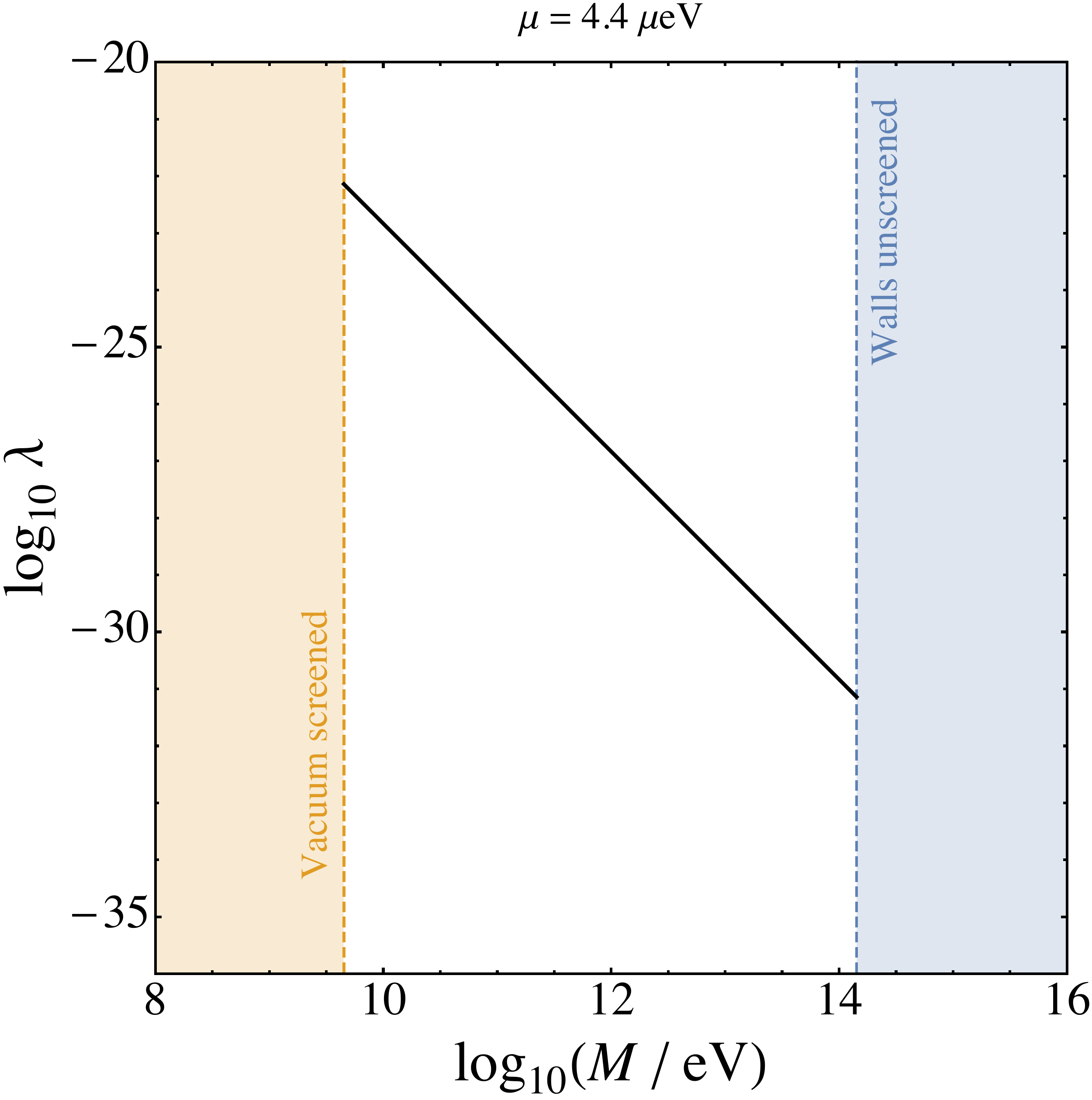}
    \caption{\scriptsize Symmetron parameters that eliminate the tension between experimental and theoretical values of $a_\mu$, for $\mu = R_\mathrm{vac}^{-1} = 4 ~\mu \mathrm{eV}$. The calculation of the classical effect is only valid in the unshaded middle region.  The black curve is a sum of the classical and quantum contributions to the anomalous magnetic moment, although the classical component is the dominant one in this regime, while the quantum component dominates for $M \lesssim 10^{-2}~\mathrm{GeV}$.  This plot is for illustrative purposes only, as at this mass the bouncing neutron experiment~\cite{Jenke:2020obe} rules out the symmetron parameter space shown here.}
    \label{fig:constraints1}
\end{figure}

Recall that we have assumed the critical density $\mu^2 M^2$ lies between the densities of the vacuum and the walls (Eq.~\eqref{crit-density}).  Using $\mu \approx R_\mathrm{vac}$, we have
\begin{equation}
    R_\mathrm{vac} \sqrt{\rho_\mathrm{vac}} \ll M \ll R_\mathrm{vac} \sqrt{\rho_\mathrm{wall}}~.
\end{equation}
The quantum contribution is given by Eq.~\eqref{quantum-contribution}
\begin{equation}
    \delta a_\mathrm{QM} = 3 \left( \frac{m_\mu}{4 \pi} \right)^2 \frac{\mu^2}{M^4 \lambda}~.
    \label{symm-qm}
\end{equation}
In Fig. \ref{fig:constraints1} we have plotted in the $(M,\lambda)$ plane the region of parameter space where symmetrons of mass $\mu \approx R_{\rm vac}^{-1}$ could account for the anomalous magnetic moment of the muon.

For  $\mu \sim 10^{-5}$ eV  as specified by the size of the storage ring, quantum neutron experiments \cite{Jenke:2020obe} exclude all models with $\lambda \le 10^{-20}$ and $M\le 10^{2}$ TeV. This excludes the large $M$ region of the parameter space where the classical contribution to the anomalous spin precession is dominant.  As such, the classical effects of a symmetron model cannot account for the observed discrepancy between experiment and theory of the anomalous magnetic moment.

We now turn to the first scenario listed above where $\mu^{-1} \ll R_\mathrm{vac}$. In this regime the classical effect is zero but the quantum contribution could still be significant.   Since we will consider symmetrons with much larger masses $\mu$, we restore the function $I_1$ and seek symmetron parameter values that solve the equation
\begin{equation}
    \delta a_\mu = 2 \left( \frac{m_\mu}{4 \pi} \right)^2 \frac{\mu^2}{M^4 \lambda} I_1 \left( \frac{\sqrt 2 \mu}{m_\mu} \right)~.
    \label{symm-qm-only}
\end{equation}
We still require that the vacuum density $\rho_\mathrm{vac}$ is sufficiently small for the symmetron to have a non-zero vev, so we have
\begin{equation}
    \rho_\mathrm{vac} \ll \mu^2 M^2~,
    \label{high-vacuum}
\end{equation}
as before.  However, we no longer require that the vacuum chamber walls be sufficiently dense, as the field remains near its vev $v \approx \mu / \sqrt{\lambda}$ everywhere.

\begin{figure}[t]
    \centering
    \includegraphics[width=0.5\textwidth]{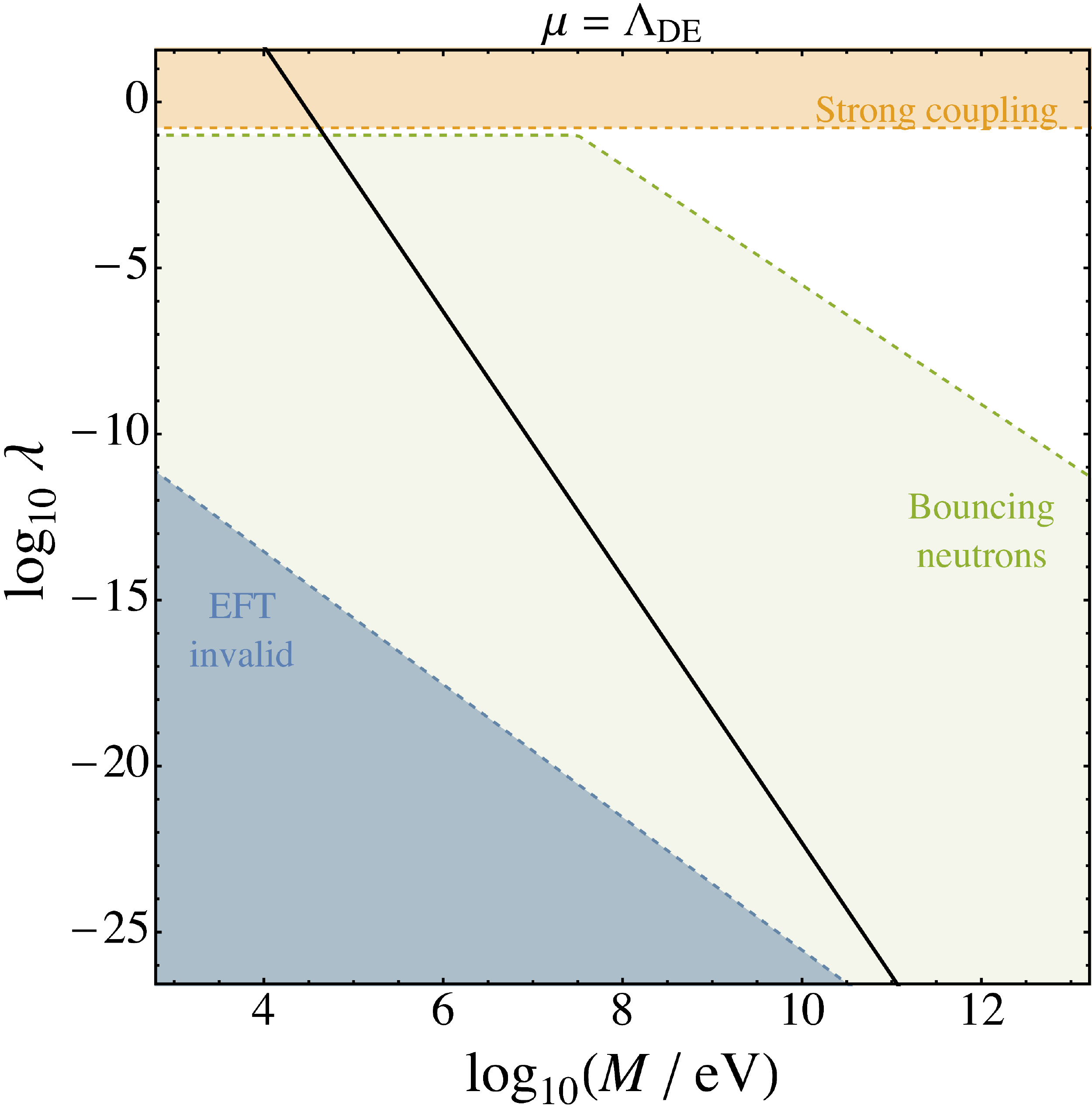}
    \caption{\scriptsize Symmetron parameters, with the mass at the dark energy scale $\mu = 2.4 ~\mathrm{meV} = \Lambda_\mathrm{DE}$, that could explain the muon $g - 2$ anomaly.  We see that all viable parameter combinations at this mass, indicated by the black line, are excluded by cold neutrons~\cite{Jenke:2020obe}.}
    \label{fig:symm-meV}
\end{figure}

Another bound on the validity of the calculation comes from the sizes of the couplings.  The symmetron introduces the non-renormalizable coupling ${\cal L}_\mathrm{int} \sim \frac{\phi^2}{M^2} \rho_m$.  There is nothing preventing higher order terms that still respect the $\phi \to - \phi$ symmetry from appearing, such as $\frac{\phi^4}{M^4} \rho_m$, $\frac{\phi^6}{M^6} \rho_m$, and so on.  As we only include the first term in this calculation, we require that the higher-order terms are suppressed, that is, $\phi^2 / M^2 \ll 1$.  The largest that $\phi$ can reach in the experiment is the vev $\phi = \mu / \sqrt{\lambda}$, giving the constraint~\cite{Brax:2018zfb}
\begin{equation}
    \frac{\mu^2}{2 \lambda M^2} \lesssim 1~.
    \label{eft-constraint}
\end{equation}

\begin{figure}[t]
    \centering
    \subfigure[]{
    \includegraphics[width=0.4\textwidth]{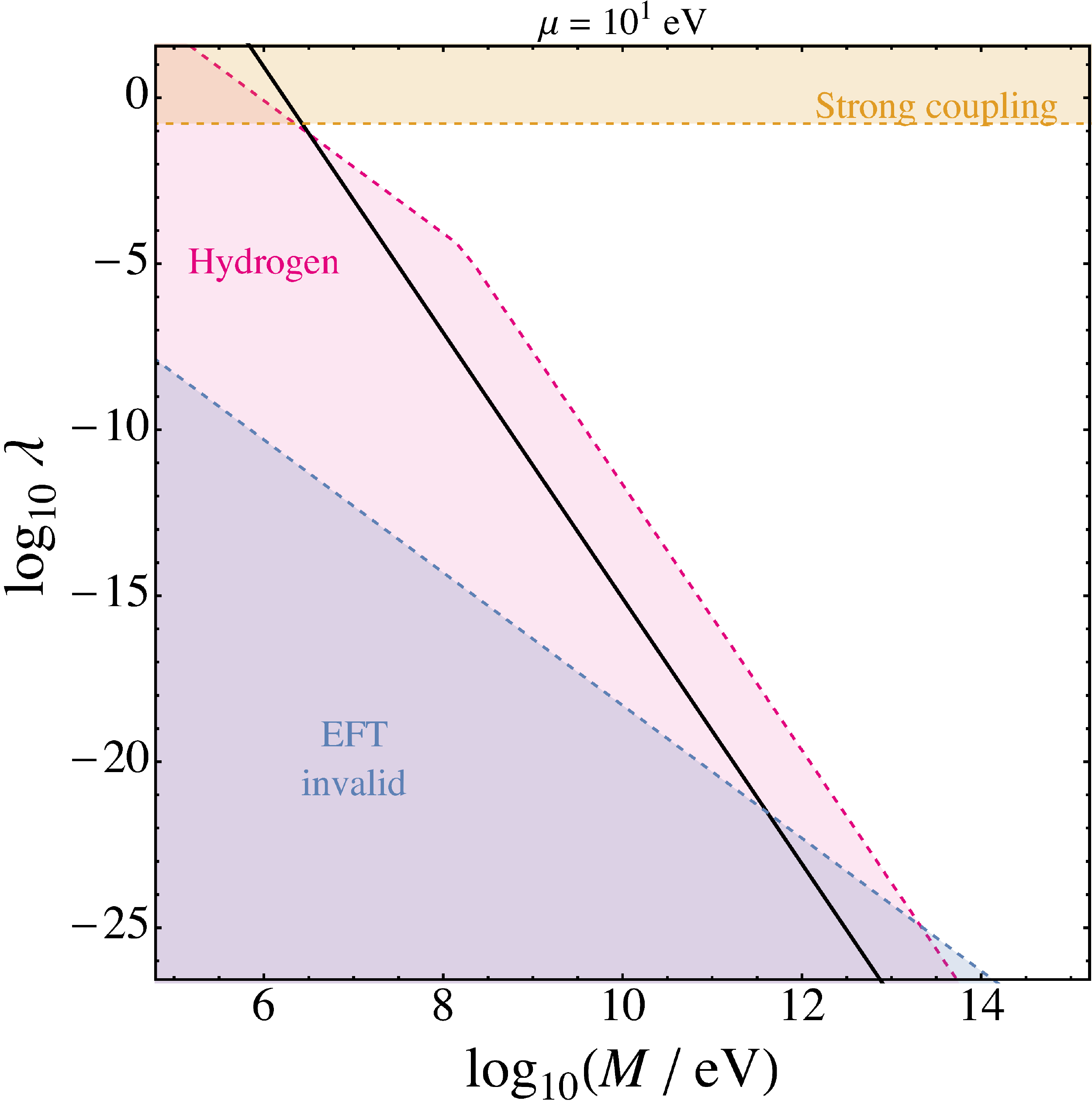}
    \label{fig:symm-1}
    }
    \subfigure[]{
    \includegraphics[width=0.4\textwidth]{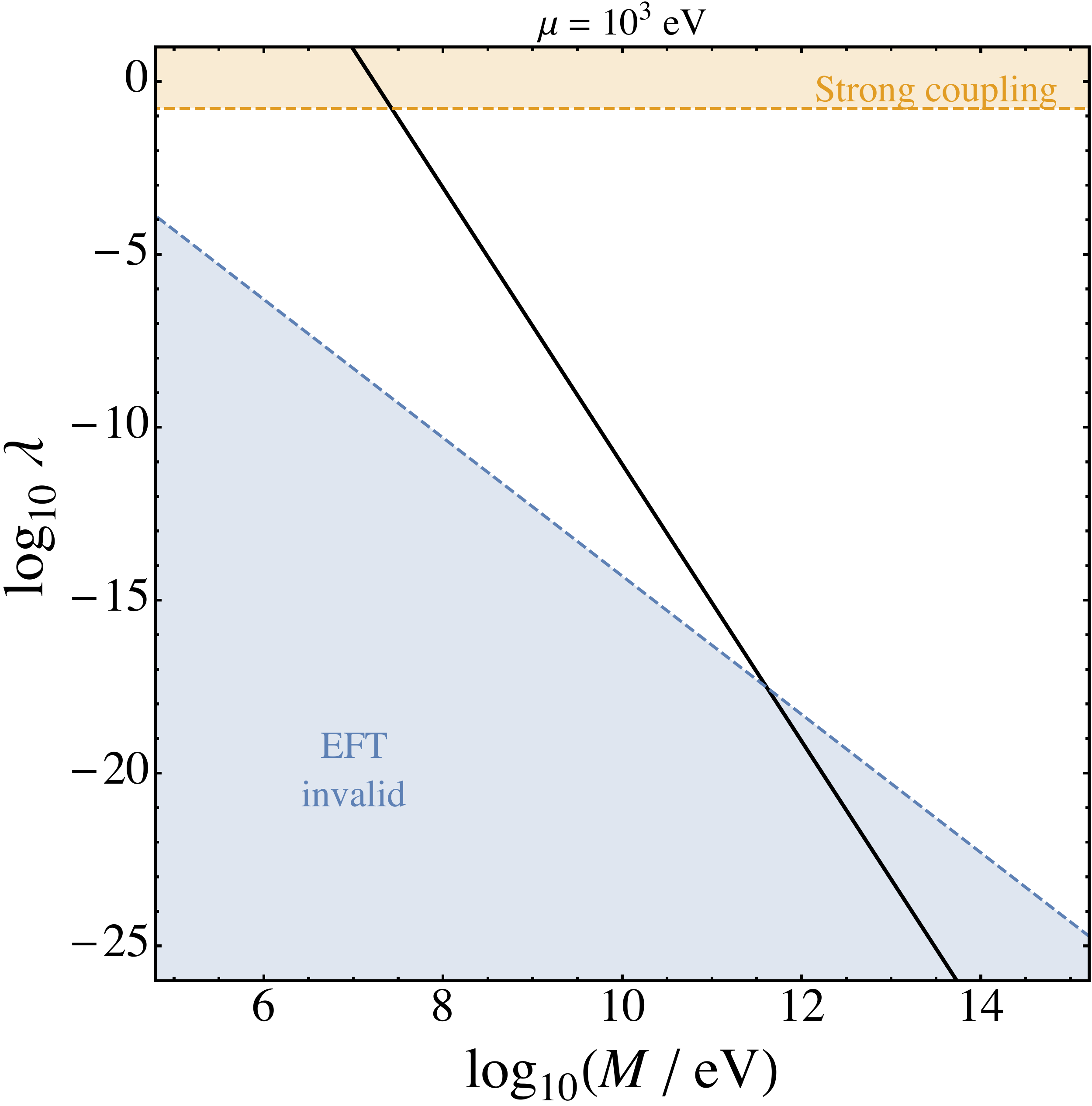}
    \label{fig:symm-2}
    }
    \subfigure[]{
    \includegraphics[width=0.4\textwidth]{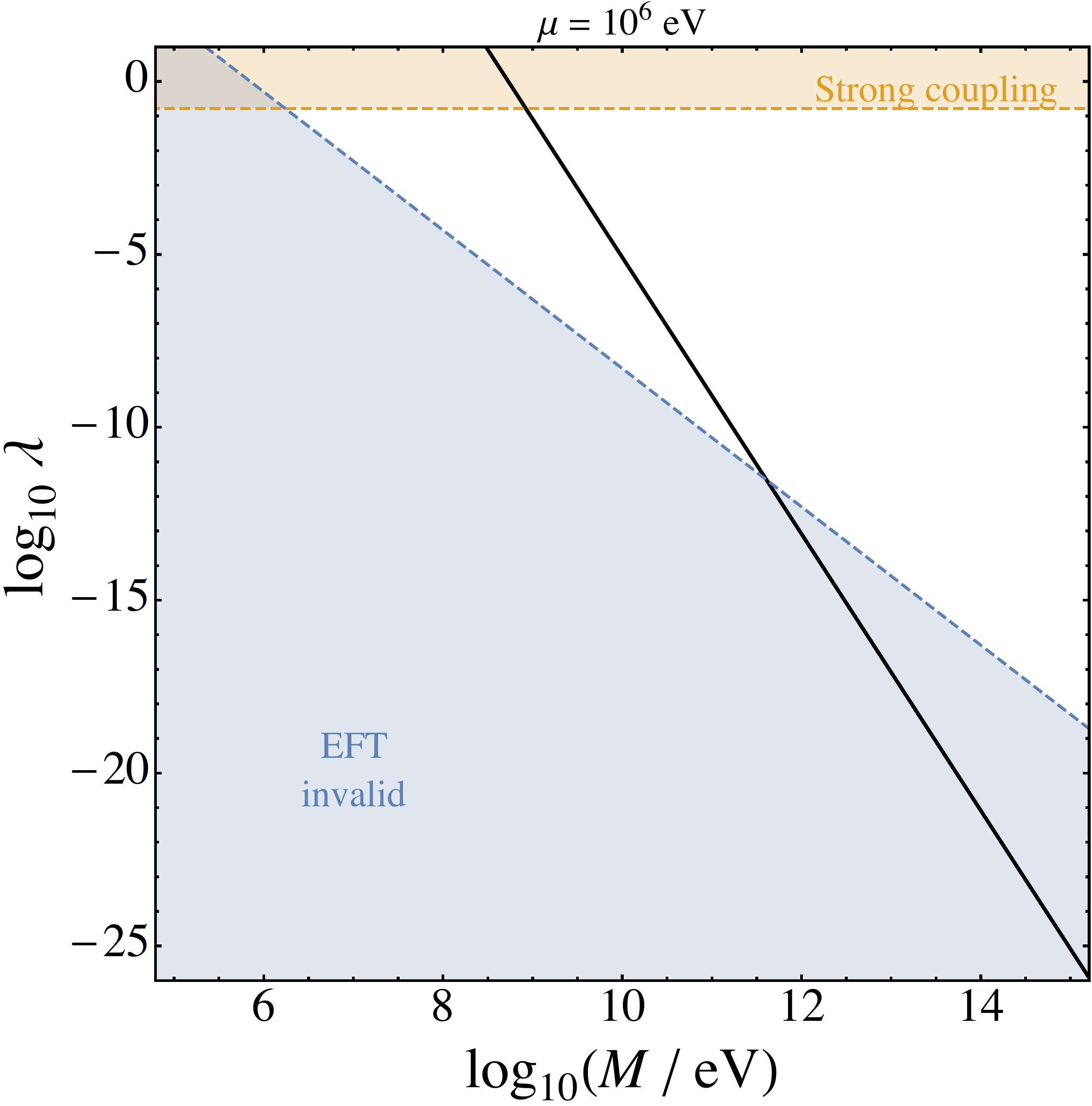}
    \label{fig:symm-3}
    }
    \subfigure[]{
    \includegraphics[width=0.4\textwidth]{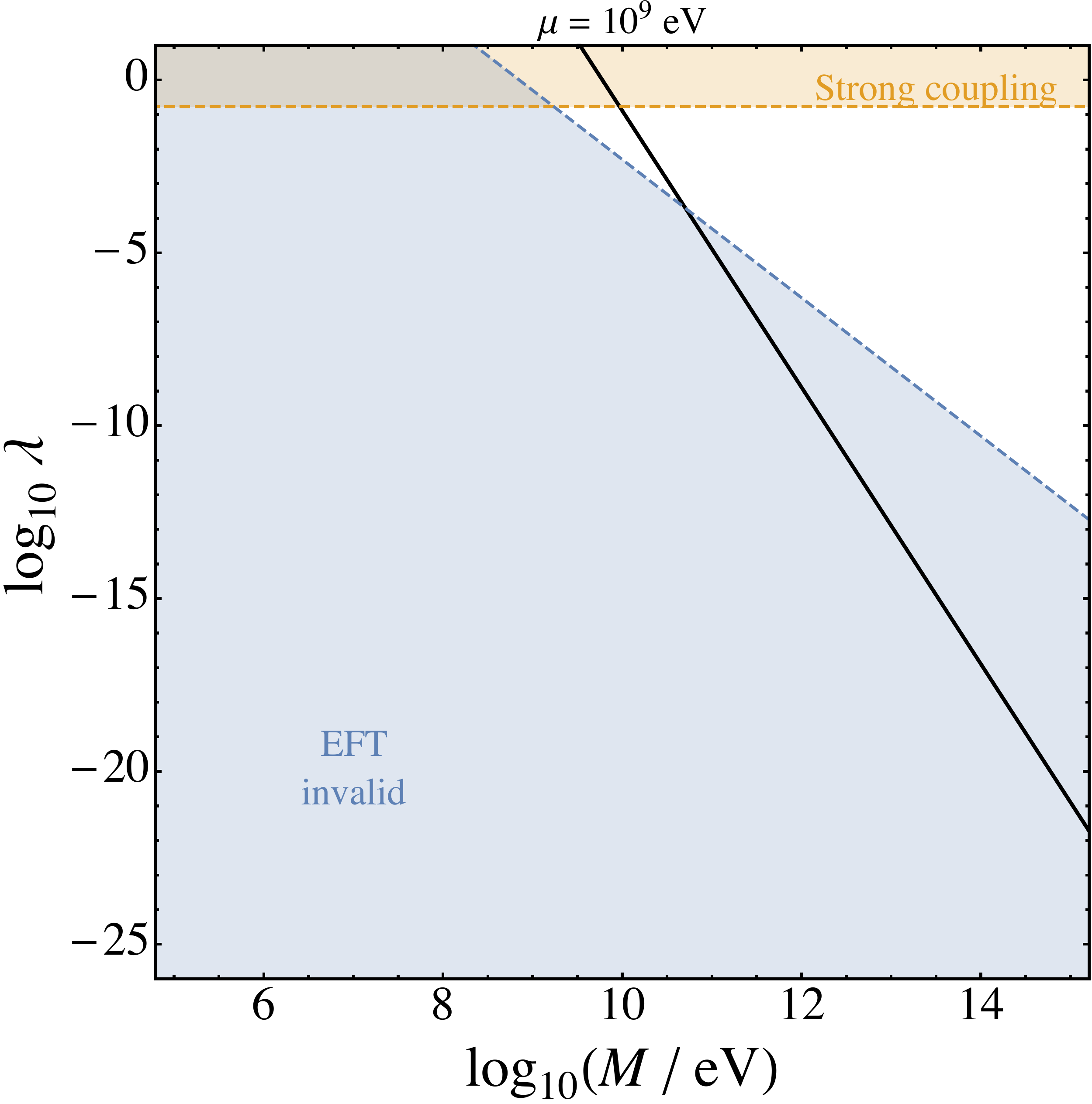}
    \label{fig:symm-4}
    }
    \caption{\scriptsize Symmetron parameters that could resolve the muon g - 2 anomaly. The combinations of parameters $\mu, M, \lambda$ that do this are indicated by the black line.  Perturbation theory breaks down for $\lambda \gtrsim 1/6$, so models in that regime are not viable.  Likewise, the EFT is not valid for very small values of $\lambda$.  Parameter combinations with $\mu \lesssim 1~\mathrm{eV}$ are ruled out by bouncing neutrons~\cite{Jenke:2020obe} and hydrogen.  Those with masses larger than $\sim 10~\mathrm{GeV}$ do not have any region where the calculation may be trusted and where the resulting correction to $\delta a_\mu$ is the right size to explain the anomaly.  Only the most limiting constraints on symmetron parameter space are plotted, corresponding to Eqs.~\eqref{eft-constraint} and~\eqref{PT-self-coupling}, { as well as new bounds deriving from precision hydrogen spectra~\cite{Brax:2022olf}.}}
    \label{fig:symmetron}
\end{figure}

The calculation of the quantum effect Eq.~\eqref{quantum-contribution} relied on perturbation theory, necessitating that the couplings are sufficiently small.  After expanding the theory (Eq.~\eqref{eq:Dirac}) about the vev, the coupling between the muons and the scalar quantum fluctuations $\chi$ are given by Eq.~\eqref{eq:L_linear_couplings}.  This depends on the dimensionless coupling strength $\beta_\mathrm{m} m_\mu / m_\mathrm{pl}$.  The magnitude of this coupling must be smaller than 1 for perturbation theory to apply, giving
\begin{equation}
    \frac{\mu m_\mu}{\sqrt{\lambda} M^2} \lesssim 1~.
    \label{PT-matter-coupling}
\end{equation}
Similarly, the scalar field's self coupling must also be sufficiently small for perturbation theory to work:
\begin{equation}
    \lambda \lesssim \frac{1}{6}~.
    \label{PT-self-coupling}
\end{equation}
All inequalities from Eq.~\eqref{high-vacuum} - \eqref{PT-self-coupling} must hold in order for the calculation of Eq.~\eqref{symm-qm-only} to be trusted.

There are three experiments to consider in this mass range.  Bouncing neutrons~\cite{Jenke:2020obe} constrain symmetrons at $1~\mathrm{eV}$ with coupling $\lambda < 1$, ruling out any possibility for smaller symmetron masses to resolve the $\delta a_\mu$ discrepancy.  This experiment, in particular, excludes the possibility for symmetrons at the same mass scale as dark energy, $\mu \approx \mathrm{meV}$, to explain the $g - 2$ discrepancy.  This is illustrated in Fig.~\ref{fig:symm-meV}.  We therefore focus on symmetrons with mass $\mu > 1~\mathrm{eV}$.  There also exist bounds from measurements of the electron magnetic moment~\cite{Brax:2018zfb}, which take a nearly identical form to Eq.~\eqref{symm-qm-only}:
\begin{equation}
    \delta a_e = 2 \left( \frac{m_e}{4 \pi} \right)^2 \frac{\mu^2}{M^4 \lambda} I_1 \left( \frac{ \sqrt 2 \mu}{m_e} \right)~,
\end{equation}
where $\delta a_e = 0.77 \times 10^{-12}$ is the uncertainty on the electron magnetic moment in the experiments~\cite{PhysRevLett.100.120801, Hanneke:2010au}.  This does not place any limitation on the parameter combinations considered here.  However, an improvement of the uncertainty on the electron magnetic moment by approximately 2 orders of magnitude would be sufficient to detect or rule out the symmetron models we discuss.  {  The third experiment to consider is precision atomic spectrometry~\cite{Brax:2010gp,Burrage:2016bwy}.  The presence of a fifth force  perturbs the energy levels of an atom, so it is possible to constrain forces with a range of approximately a Bohr radius and larger.  This translates to bounds on symmetron models that have mass $\mu \lesssim \mathrm{keV}$~\cite{Brax:2022olf}.}  There are indeed combinations of symmetron parameters $\mu, M, \lambda$ that satisfy all of these requirements and produce the right size of correction to the anomalous magnetic moment $\delta a_\mu$.  They lie in the range $1~\mathrm{keV} < \mu <  10~\mathrm{GeV}$, and are illustrated in Fig.~\ref{fig:symmetron}.

Given that the symmetron was originally motivated by, and associated with, dark sector physics~\cite{Hinterbichler:2010es, Hinterbichler:2011ca} it is natural to wonder whether any of the parameter combinations indicated in Fig.~\ref{fig:symmetron} could result in the same scales as dark energy.  We find that they cannot, so the models found here lie squarely in the realm of modified gravity rather than dark energy.  The criteria by which this determination was made are given in Appendix~\ref{app:DE-vs-MG}.

We have found that symmetron models in a wide mass range could account for the muon g - 2 anomaly.  This results from their quantum contribution at one loop, and opens up the exciting possibility that modified gravity could account for the deviation in the anomalous magnetic moment of the muon. Currently the parameter range we have uncovered is unconstrained by any other experiments. On the other hand if there were improved precision in the anomalous magnetic moment of the electron, or in precision Hydrogen experiments, by a couple of orders of magnitude our symmetron regime could either be confirmed or found to be in tension. 

\section{Discussion}

\subsection{Cosmological consequences}

The chameleon and symmetron models discussed in this paper, which could play a role in the $(g-2)_\mu$ results, are not traditional. They are characterised by parameters which are not within the range discussed at length in cosmological applications of these models. For this reason, it is of interest to survey their possible consequences for cosmological observations. We will also mention some severe constraints from particle physics which could jeopardise the validity of some of these models.

Let us start with the inverse power law chameleons. Their cosmology was discussed  in the original cosmological chameleon paper \cite{Brax:2004qh}. Let us first consider the theory at high energy in the radiation era. We must assume that the chameleon field sits at the minimum of its effective potential
\be 
\phi_{\rm cosmo}= \left(\frac{M\Lambda^5}{\rho} \right)^{1/2}~,
\ee
where we focus on the $n=1$ model here. 
The field must follow the minimum of the effective scalar potential since Big Bang Nucleosynthesis (BBN) which takes place at  a redshift of order  $z_{\rm BBN}\sim 10^9$. The mass of the field at the minimum is given by
\be 
m_\phi^2 = 2 \Lambda^5 \left( \frac{\rho}{M\Lambda^5} \right)^{3/2}~,
\ee
which grows with the density. The mass  must satisfy $m_\phi\gtrsim H$, i.e. the scalar mass must be larger than the Hubble rate, for the minimum to be stable, i.e. otherwise the minimum is not a tracking solution and the field does not follow the time evolution of the minimum \cite{Brax:2004qh,Brax:2012gr}. Let us focus on $M=10^{-16} M_{\rm Pl}$ and $\Lambda=10^{-4}$ eV as a typical value in the allowed interval $10^{-7}~\mathrm{eV} \lesssim  \Lambda \lesssim 10^{-3}~\mathrm{eV}$.  In this case we find that the mass at BBN is of the order $0.6 $ GeV, i.e. larger than the typical temperature of $T\sim 1 $ MeV, and the chameleons are non-relativistic. At higher temperature, as $m_\phi \propto a^{-9/4} \propto T^{9/4}$ since the chameleon mass depends on the matter energy density varying in $a^{-3}$, they are even more non-relativistic and cannot be produced thermally as the typical energy $E\simeq T \ll m_\phi$.  
The time variation of the chameleons since BBN would induce a variation of particle masses~\cite{Brax:2004qh}
\be 
\frac{\Delta m}{m}= \frac{\vert \Delta \phi \vert}{M}~,
\ee
which is dominated by the value of $\phi$ in the cosmological background $\phi_0$. For the chosen parameters, we find a variation of order $6 \times 10^{-11}$ which is completely negligible. 
Later in the matter era, the chameleon becomes lighter and could modify two crucial observables. First of all, if the chameleon energy density varies in time, the equation of state of dark energy would vary and this would affect the angular distance to the last scattering surface and therefore would shift the peaks of the Cosmic Microwave Background (CMB). The $\phi$ dependent part of the chameleon potential $\frac{\Lambda^5}{\phi}$ evaluated at $z_{\rm CMB} \sim 10^3$ would be $10^{-5}$ the amount of dark energy, i.e. the field dependent part of the chameleon potential is negligible between last scattering and now. Similarly the kinetic terms are given by $\frac{9H^2}{8} \phi^2$ which is negligible in the Friedmann equation as $\phi\ll M_{\rm Pl}$. Hence the field dependent part of the chameleon energy density leads to a minute deviation of the dark energy equation of state from $\omega_{\rm DE}=-1$.  On the other hand, the chameleon could affect the growth of structure. Indeed, structure grows with an enhancement factor of $(1+ 2 \frac{M_{\rm Pl}^2}{M^2})$  inside the Compton wavelength of the scalar field \cite{Brax:2004qh,Brax:2021wcv}. This is valid in linear perturbation theory and can be trusted as long the Compton wavelength is larger than a few Mpc's. If the Compton wavelength of the scalar field is smaller then the effects of the scalar field are blurred by astrophysical effects on galactic scales and below. As the enhancement factor in our model is large, one must require that the Compton wavelength should be sufficiently small. For our template, $m_\phi \sim 4\times 10^{-21}$ GeV at the present time, which is twenty orders of magnitude larger than the Hubble rate, i.e. the Compton wavelength is much smaller than cosmological and astrophysical scales. In conclusion, we find that the chameleon which could play a role for $(g-2)_\mu$ has no impact on cosmology. 
Finally let us comment briefly on what happens for symmetrons. As shown in Appendix~\ref{app:DE-vs-MG}, we do not expect the symmetrons described here to have any cosmological consequences for values of $\mu\gtrsim 1$~MeV.

\subsection{Coupling to the Higgs}

The chameleon could also have consequences for particle physics experiments. In particular, the chameleon could lead to new channels in the decay of the Higgs particle. This invisible Higgs decay is constrained by the branching ratio $BR(H\to inv)<11 \%$ by the LHC results at run2 \cite{Arcadi:2021mag}. The decay of the Higgs field into a pair of chameleons  would follow from the direct coupling of the chameleon to the Higgs as
\be 
\delta {\cal L}_{H\phi}= {\mu_H^2 A_H^2(\phi)} H^\dagger H - {\lambda_H }(H^\dagger H)^2~,
\label{potH}
\ee
 where the coupling function $A_H(\phi)$ is absent from the quartic self-interaction. The scaling in $A_H(\phi)$ corresponds to the dimension of the coupling constant. Here we assume that the Higgs field couples to the rescaled metric $g^H_{\mu\nu}= A^2_H(\phi) g_{\mu\nu}$. We distinguish $A_H$ from the coupling function $A$ to fermions as they play very different physical roles as we will see below.  For the complex scalar field $H_E$ coupled to $g_{\mu\nu}^H$, the Higgs-scalar  Lagrangian reads 
\be 
{\cal L}_H=-A_H^2(\phi){\partial_\mu H^\dagger_E \partial^\mu H_E} + A_H^4(\phi)\mu_H^2 H^\dagger_E H_E - A_H^4(\phi)\lambda_H (H^\dagger_E H_E)^2~,
\ee
where the indices are raised using the metric $g_{\mu\nu}^H$. Normalising the field as $H= A_H(\phi) H_E$ and focusing on the non-derivative interaction terms leads to Eq.~\eqref{potH} and the Lagrangian
\begin{equation}
    {\cal L}_H = - \partial H^\dagger \partial H + \mu_H^2 A_H^2 H^\dagger H - \lambda_H (H^\dagger H)^2~.
    \label{Higgs-scalar-unbroken}
\end{equation}
As we are interested in the effects of the real Higgs boson after electroweak symmetry breaking, let us decompose 
\begin{equation}
    H = \frac{1}{\sqrt 2} \left( \begin{matrix}
    0 \\
    \varphi_H
    \end{matrix}
    \right)~,
\end{equation}
so that the Lagrangian becomes
\begin{equation}
    {\cal L}_H = - \frac{1}{2} (\partial \varphi_H)^2 + \frac{1}{2} \mu_H^2 A_H^2 \varphi_H^2 - \frac{\lambda_H}{4} \varphi_H^4~.
\end{equation}
Since $A_H(\phi)$ will be expanded around unity (i.e. the absence of coupling) it is convenient to isolate the Higgs-scalar interaction operator
\begin{equation}
    {\cal L}_H = - \frac{1}{2} (\partial \varphi_H)^2 + \frac{1}{2} \mu_H^2 \varphi_H^2 - \frac{\lambda_H}{4} \varphi_H^4 + \frac{1}{2} \mu^2 (A_H^2 - 1) \varphi_H^2~.
\end{equation}
When matter is present with a density $\rho$, the vev $\langle \phi \rangle$ is determined by 
\be 
\frac{\partial V}{\partial \phi}= -\frac{\mu_H^2 \varphi^2_H}{2} \partial_\phi A^2_H(\phi)-\partial_\phi A \rho~,
\label{min1}
\ee
where the second term is responsible for the environmental dependence of the vev $\langle \phi \rangle (\rho)$.

In the spirit of effective theories, we assume that in fact we can expand $A_H(\phi)$ in powers of $\phi/M$ such as 
\be 
A_H^2(\phi)= 1 + \frac{2\kappa_\phi\phi}{M} + \lambda_\phi \frac{\phi^2}{M^2}+ \dots
\label{AH-coupling}
\ee
The first two terms reconstruct $A(\phi) \simeq 1+\frac{\phi}{M}$ when $\kappa_\phi=1$. This is the function used in this paper for chameleons for the  coupling  between the scalar field and fermions. For symmetrons we have $\kappa_\phi=0$ and $\lambda_\phi=1$. When $\kappa_\phi$ and $\lambda_\phi$ differ from these values, we have $A_H\ne A$ and the Higgs boson coupling is not equal to the fermion coupling function. 

The Higgs vev is determined by
\be 
\lambda_H \varphi_H^2= A_H^2(\phi) \mu_H^2~.
\label{min2}
\ee
In practice we assume that  $A_H\sim 1$  and we retrieve
the Higgs vev $ v = \mu_H / \sqrt{\lambda_H}$ as usual. 
Expanding $\varphi_H = v + h $ and $\phi= \langle \phi \rangle (\rho) + \delta\phi$, we find for the Higgs part of the Lagrangian
\begin{equation}
    {\cal L}_H = - \frac{1}{2} (\partial h)^2 - \mu_H^2 h^2 + \frac{1}{2}\mu_H^2 (A_H^2 - 1) (v^2 + 2 v h + h^2)~,
    \label{higgs-lagr}
\end{equation}
complemented with the Higgs cubic and quartic self-couplings which are not needed for the present discussion.  At leading order and neglecting the coupling to the scalar $\delta\phi$ we can now identify the Higgs mass
\begin{equation}
    m_h = \sqrt 2 \mu_H~.
\end{equation}
Expanding around $\langle \phi\rangle$ we have
\be 
A^2= 1 + \frac{2\kappa_\phi\langle \phi\rangle }{M} + \lambda_\phi \frac{\langle \phi\rangle ^2}{M^2} + \frac{2}{M}\left(\kappa_\phi + \lambda_\phi \frac{\langle \phi\rangle}{M}\right) \delta \phi + \lambda_\phi \frac{\delta\phi^2}{M^2}+\dots
\ee
where the constant term is very close to unity. At quadratic order, the Higgs-scalar Lagrangian becomes
\be 
{\cal L}_{h-\delta\phi}\supset - \frac{1}{2} (\partial h)^2 - \mu_H^2 h^2 +\left(\lambda_\phi\frac{\mu_H^2 v^2}{M^2}- V''-A''\rho \right) \frac{\delta\phi^2}{2} + \frac{vm^2_h}{M}\left(\kappa_\phi + \lambda_\phi \frac{\langle \phi\rangle}{M}\right) h\delta \phi~.
\ee
Notice that the terms linear in $\delta \varphi$ and $h$ cancel thanks to the minimum equations Eqs.~\eqref{min1} and \eqref{min2}. As we can see the mass matrix is not diagonal and the Higgs field $h$ is no longer a mass eigenstate:
\be 
{\cal M}^2_{}=\left (
\begin{array} {cc}
m^2_h & -\frac{vm^2_h}{2M}(\kappa_\phi + \lambda_\phi \frac{\langle \phi\rangle}{M})\\
-\frac{vm^2_h}{2M}(\kappa_\phi + \lambda_\phi \frac{\langle \phi\rangle}{M})&  
 m^2_\phi- \lambda_\phi \frac{\mu_H^2 v^2}{M^2}\\
\end{array}
\right )~.
\ee 
Nothing guarantees now that one of the eigenmodes is a light scalar field of mass
\be 
m^2_\phi= V'' + A''\rho~,
\ee
anymore. Here we denote by  $'=\frac{d}{d\phi}$ evaluated at the minimum $\langle \phi\rangle (\rho)$.

Setting aside this problem for now, we focus on the next term in the EFT, i.e.
the cubic $h\delta \phi^2$ interaction term
\begin{equation}
    {\cal L}_{h\delta\phi} =  \lambda_\phi \frac{v m_h^2}{2M^2} \delta\phi^2 h~.
\end{equation}
This term can be interpreted as springing from an effective quartic interaction between the Higgs doublet and the scalar field such as \cite{Arcadi:2021mag}
\begin{equation}
    {\cal L}_4 \supset  \frac{1}{4} \lambda_{H\phi\phi} H^\dagger H \phi^2~,
    \label{Higgs-scalar-quadratic}
\end{equation}
which leads us to identify the quartic coupling as 
\begin{equation}
    \lambda_{H\phi\phi} = 2 \lambda_\phi \frac{m_h^2}{M^2}~.
\end{equation}
For a scalar with mass $\lesssim 80$~GeV, the Higgs-scalar coupling is constrained to be $\lambda_{H\phi\phi} < 0.01$ from Higgs boson decay.  This translates into a bound
\begin{equation}
    M > 10 \sqrt{2\lambda_\phi} m_h \approx ~\sqrt{\lambda_\phi}\ \mathrm{TeV}~.
\end{equation}
This would appear to rule out all of the chameleon and symmetron $g - 2$ models if $\lambda_\phi={\cal O}(1)$.  

In fact, this phenomenological constraint is superseded by bounds from the existence and validity of screening in the presence of a Higgs-scalar coupling. To see this, let us come back to the mass matrix of the Higgs-scalar system.
The trace of the mass matrix gives the sum of the two square eigenmasses and we should request that this trace be only weakly perturbed to guarantee that both the stability of the electroweak symmetry breaking and that the scalar mass remains small compared to the electroweak scale.
We can see that Eq.~\eqref{Higgs-scalar-quadratic} contributes to the scalar mass element ${\cal M}^2_{\phi\phi}$ by an amount
$
    \lambda_{H \phi \phi}v^2/2~.
$
We need this to be smaller than the  scalar mass squared  in the vacuum regime of the $g - 2$ measurement, i.e.
\be 
\lambda_{H\phi\phi} \ll \frac{m^2_\phi}{v^2}~.
\ee
As the vacuum chamber has a size of order 10~cm, this translates to a minimum mass of $10^{-6}~\mathrm{eV}$ for the scalar, yielding a maximum Higgs-scalar coupling of $\lambda_{H\phi\phi} \lesssim 10^{-34}$. This is obviously much more stringent than the Higgs decay constraint and implies also that 
\be 
\lambda_\phi \lesssim 10^{-34}~,
\ee
when $M\simeq 10^{-16} M$ as is the case for chameleon models reproducing $(g-2)_\mu$.  This reasoning extends straightforwardly to arrive at similar conclusions for symmetrons. 

Next, let us  consider what happens in the presence of matter to the minimum of the effective potential $\langle \phi \rangle (\rho)$.  Theories like the chameleon and symmetron rely sensitively on variations in the ambient energy density.  But, it can be seen from Eqs.~\eqref{AH-coupling} and \eqref{higgs-lagr} that the scalar field is also sensitive to the Higgs condensate energy:
\begin{equation}
    {\cal L}_\phi \supset -\kappa_\phi \frac{\mu_H^2 v^2}{M} \phi - \frac{\rho}{M} \phi~,
\end{equation}
This energy is enormous compared to typical densities in the laboratory, and threatens to overwhelm the chameleon screening dynamics.  The only resolution is to fine-tune the Higgs coupling parameter $\kappa_\phi$ to a small value.  It follows that the Higgs coupling does not destabilise the scalar vacuum provided
\be 
\kappa_\phi \ll \frac{\vert M A' \rho \vert}{2 \mu^2_H v^2}~.
\ee
Assuming $A'(\phi) = M^{-1}$, as is the case with the chameleon, and using the ambient matter density $\rho_\mathrm{vac} \approx 10^6~\mathrm{eV}^4$ in the muon $g - 2$ experiment, this translates into a bound
\begin{equation}
    \kappa_\phi \lesssim 10^{-38}~,
\end{equation}
for chameleons and an even more stringent one for symmetrons.

We can see that keeping our scalar field light, and insensitive to the Higgs condensate energy, requires us to fine tune the Higgs coupling to be so small that the constraints from invisible Higgs decays are automatically satisfied.
In practice, this tells us that although $A(\phi)\ne 1$ to reproduce $(g-2)_\mu$, we must impose that $A_H(\phi)=1$ to preserve the screening mechanisms for both chameleons and symmetrons. Of course, the decoupling of the screened scalars from the Higgs boson sector will be challenged by quantum corrections in a way reminiscent of the hierarchy problem for the Higgs boson itself. A detailed analysis of this issue is left for future work.

\subsection{Other Particle physics effects}

Ruling out the chameleon or symmetron explanation of $(g-2)_\mu$ completely would require additional experimental input. This may be provided by the physics of $B$ mesons, as the coupling of the scalar to fermions could lead to interesting effects in $B$ meson decays \cite{Graverini:2018riw}. In particular, the scalar with a linear coupling to matter coupling to fermions is proportional to the mass of the fermions \footnote{The coupling $\phi \rho \sim \phi m_\psi \bar \psi \psi$ assumes that the fermions are non-relativistic, which is certainly not always the case in $B$ decays.  The relativistic effects are not considered in the following but should be reinstated to obtain accurate bounds on the modified gravity parameters.}
\begin{equation}
    {\cal L}_\mathrm{int} = -\frac{\phi}{M} \rho = -\sum_\psi \frac{\phi}{M} m_\psi \bar \psi \psi~,
    \label{B-linear-coupling}
\end{equation}
where the sum runs over all Standard Model fermion fields $\psi$.  This interaction can therefore potentially break the universality of $B$ meson decays \cite{LHCb:2017rln}. This would contribute to the anomalous behaviour of these decays and lead to a ratio
$R_K= \frac{BR(B\to K\mu\mu)}{BR(B\to Kee)}$ which would differ from unity. As an order of magnitude estimate from penguin diagrams where an intermediate photon is replaced by a scalar,  we expect
\be 
\frac{\Delta R_K}{R_{\rm K,SM}}\simeq \frac{m_t m_\mu}{\alpha M^2}~,
\label{B-meson-branching}
\ee
for the deviation due to scalars compared to the SM model value.
Notice that numerical factors such as $4\pi$'s have not been included in this simple estimate. This has potential consequences for the viability of our scenario especially for symmetrons. A more accurate calculation should be performed to confirm the results below ~\cite{B-paper}.
We have also neglected the mass of the scalar.  This could also be important for symmetrons with masses $\gtrsim~\mathrm{MeV}$ and would suppress the contribution due to these scalars.

Imposing that the ratio in Eq.~\eqref{B-meson-branching} should be less than $10^{-2}$ leads to $M\gtrsim 150~\mathrm{GeV} = 10^{-16} \Mpl$.  The chameleon is already equipped with a linear coupling like Eq.~\eqref{B-linear-coupling}, and we can see from Fig.~\ref{fig:chameleon} that this criterion is marginally satisfied.

The symmetron Lagrangian, on the other hand, does not have the same linear coupling, as its leading coupling to matter is quadratic.  The linear coupling appears only after we perturb the Lagrangian about the symmetron vev $\phi = \mu / \sqrt{\lambda} + \varphi$.  The interaction term becomes
\begin{equation}
    {\cal L_\mathrm{int}} \sim -\sum_\psi \frac{\mu}{\sqrt{\lambda} M^2} \phi m_\psi \bar \psi \psi~.
\end{equation}
Making the replacement $\frac{1}{M} \to \frac{\mu}{\sqrt \lambda M^2}$ in Eq.~\eqref{B-meson-branching}, we find
\begin{equation}
\frac{\Delta R_K}{R_{\rm K,SM}}\simeq \left( \frac{\mu}{\sqrt{\lambda} M^2} \right)^2 \frac{m_t m_\mu}{\alpha}~.
\end{equation}
Once again, we impose that the ratio $\frac{\Delta R_K}{R_{\rm K,SM}}$ must be less than $10^{-2}$, and obtain the constraint
\begin{equation}
    \frac{\mu}{\sqrt{\lambda} M^2} < 2.0 \times 10^{-3} ~\mathrm{GeV}^{-1}~.
\end{equation}
This can be compared with Eq.~\eqref{symm-qm-only}, which shows that when the symmetron mass is negligibly much less than the muon mass (as has also been assumed here) we require
\begin{equation}
    \frac{\mu}{\sqrt{\lambda} M^2} = 3.5 \times 10^{-3} ~\mathrm{GeV}^{-1}~.
    \label{symmetron-combination-g2}
\end{equation}
for the symmetron model to account for the $g - 2$ discrepancy.  Our scenario is therefore in mild tension with B meson decay.  Of course, this order of magnitude  should be confirmed by an exact calculation that also accounts for the scalar mass and the full relativistic scalar-fermion coupling.

Strong constraints on the coupling of the scalar to fermions can also be deduced from the decay of kaons such as $K^+\to \pi^+\phi$ \cite{Beacham:2019nyx,NA62:2020xlg,NA62:2021zjw}. In particular the NA62 has put bounds on the interactions between a scalar particle and matter.  That bound is given within the context of a Higgs-mediated interaction between the scalar and SM particles, so we must translate that bound into the direct interaction we have here.

The relevant model for our purposes is the BC4 model~\cite{Beacham:2019nyx}, in which the scalar interacts only with the Higgs:
\begin{equation}
    {\cal L} \supset g_{\phi h} \phi H^\dagger H~.
\end{equation}
After spontaneous symmetry breaking, and integrating out the Higgs boson, there is an effective scalar-fermion interaction
\begin{equation}
    {\cal L} \supset \frac{\theta}{v} \phi \sum_\mathrm{SM} m_\psi \bar \psi \psi~,
\end{equation}
where the sum runs over all SM fermion fields, $v$ is the Higgs vev, and the mixing angle, assumed to be small, is
\begin{equation}
    \theta \approx \frac{g_{\phi h} v}{m_h^2}~.
\end{equation}
Comparing to Eq.~\eqref{B-linear-coupling}, we see that the conversion to our chameleon model parameter $M$ is
\begin{equation}
    M = \frac{v}{\theta}~.
\end{equation}
The mixing angle is experimentally constrained to be $\theta \gtrsim 10^{-4}$~\cite{NA62:2021zjw}, yielding the bound
\begin{equation}
    M > 10^6~\mathrm{GeV}~.
    \label{kaon-bound}
\end{equation}
Such  a bound is incompatible with the $(g-2)_\mu$ scale for $M$ when applied to chameleon models. For symmetrons with a mass greater than 1 MeV but smaller than 1 GeV, i.e. the symmetrons which are not excluded by cosmology and can play a role for $(g-2)_\mu$, similar conclusions can be drawn. Their effective coupling to matter is $M_\mathrm{eff} = \frac{\sqrt \lambda M^2}{\mu}$, and we see from Eq.~\eqref{symmetron-combination-g2} that $M_\mathrm{eff} = 285~\mathrm{GeV}$ is required for the theory to account for the muon $g - 2$ result, a value that is in tension with Eq.~\eqref{kaon-bound}. A more rigorous analysis is required in order to confirm that the whole symmetron parameter space depending on $(M,\mu,\lambda)$ is ruled out by kaon decays. This is left for future work.


Of course, taking into account all the phenomenology of $K$ and $B$ decays goes beyond the present analysis. Further study   should  confirm the stringent restrictions on the parameter space of chameleon and  symmetron models compatible with $(g-2)_\mu$ that we have indicated here, and a detailed analysis is currently underway~\cite{B-paper}. Throughout this paper we have insisted on the universality of the coupling of the scalar field to matter via a unique Jordan metric. This assumption can be lifted, and a model where the scalar couples most strongly to muons should favour the description of $(g-2)_\mu$ by screened modified gravity, while softening the tension with other observable such as kaon decays. This also is left for future work.

\section{ Conclusion}

In this paper we have computed the effect of a scalar field coupled to matter on the anomalous magnetic moment of the muon in the Fermilab experiment. Such a scalar field generically arises in modified gravity models. In particular we have shown that chameleon modified gravity could account for the discrepancy between the experimental and theoretical value of the anomalous magnetic moment.  We have shown that an unscreened scalar with a Yukawa coupling to matter cannot contribute significantly to $(g-2)_\mu$, and have demonstrated the powerful effects of screened modified gravity models.  The chameleon or symmetron  models could be verified or ruled out by improved electron magnetic moment experiments that are roughly two orders of magnitude more sensitive than existing ones, or by precision atomic spectroscopy experiments. Such experiments would delve further into chameleon and symmetron modified gravity parameter space and either confirm or rule out the results we have uncovered.

As part of our work we considered in detail how the classical effects of a scalar field influence the determination of the anomalous magnetic moment, hitherto not considered previously; we give a detailed analysis of the scalar field's contribution to the precession of the spin vector. This is crucial to the analysis although, for the chameleon and the symmetron models considered here the quantum contributions turn out to give the dominant effect.

We have also considered other potential constraints on our model parameters. Concentrating on chameleon models,  we showed that the chameleon mass is much larger than the typical temperature at the time of BBN, and will therefore not disturb BBN dynamics. Similarly, the chameleon models investigated here have negligible effects on the CMB and on the time variation of masses of fundamental particles. We also considered the effect the chameleon could have on several particle physics measurements.  We showed that although Higgs decays could potentially constrain the chameleon and symmetron $g - 2$ models, demanding that the Higgs condensate does not spoil the screening behavior necessitates decoupling the scalar and Higgs fields from one another.
We also briefly discussed the effect the chameleon could have on B physics, but the detailed evaluation is left to future work \cite{B-paper}. Similarly, we have also pointed out that kaon decays into pions and invisible matter are in tension with our result on $(g-2)_\mu$. This may rule out the specific chameleon and symmetron models but leaves open other avenues for screened modified gravity. For instance, another model like the environment dependent dilaton \cite{Brax:2010gi} whose screening is similar to the symmetron's should be investigated. One could also explore  a violation of universality with different coupling strengths to muons, electrons and quarks.

The BNL/Fermilab muon magnetic moment anomaly has now persisted for well over a decade.  Although the chameleon and symmetron models shown here are in tension with particle physics bounds, it is possible that a closely related model, likely with non-universal coupling to matter fields, could explain this anomaly.  It would therefore be interesting to test other models against this result. Similarly, this motivates further testing of these models in the laboratory, and raises the possibility that BNL and Fermilab may well have just discovered modified gravity.

{\small {\bf Acknowledgments}~We thank Clare Burrage, Jeremy Sakstein, Leong Khim Wong, Bipasha Chakraborty, and Ben Allanach for helpful discussions.  We particularly thank Brendan Casey for his help in understanding the experimental details. We thank M. Madigan and M. Ubiali for discussions on B meson physics. We are also grateful to the anonymous referee for their comments on potential constraints we had not considered. P.B. acknowledges support from the European Union’s Horizon 2020 research and innovation programme under the Marie Skodowska-Curie grant agreement No 860881-HIDDeN. A.C.D.  acknowledges  partial  support  from the   STFC   Consolidated   Grants   No.   ST/P000673/1 and No. ST/P000681/1.}

\appendix

\section{The angular velocity tensor}
\label{app:tenso}
The angular velocity vector can be embedded in an antisymmetric tensor $\Omega_{\mu\nu}$
such that
\be 
\frac{d u^{\mu}}{d\tau}= -\Omega^{\mu}_\nu u^\nu~,
\ee
with $\Omega_{a0}=0$ and
\be 
\omega_c^a= -\frac{1}{2\gamma} \epsilon^{abc} \Omega_{bc}~.
\ee
This is only possible as $E^av_a=0$ and $\partial_a\phi  v^a=0$. The spatial part of the antisymmetric tensor is
\be 
\Omega_{ab}=-\frac{q}{m_E}B_{ab} +E_{ab}~.
\ee
We have defined the antisymmetric tensor $B_{\mu\nu}$ by
\be
B_{ab}= \epsilon_{abc} B^c,\ B_{0a}=0~,
\ee
and the antisymmetric tensor $E_{\mu\nu}$
\be 
E_{0a}=0~,\ E_{ab}= \epsilon_{abc}{\cal E}_c~,
\ee
where we have introduced the  vector
\be
{\cal E}_a= \epsilon_{abc} v^b \left(\frac{q}{m_Ev^2}E^c -\frac{\partial_\phi \ln A}{v^2\gamma}\partial^c \phi \right)~
\ee
such that
\be 
\omega_c^a= -\frac{q}{m_E\gamma }B^a + \frac{{\cal E}^a}{\gamma}~.
\ee
The two terms vary differently under Lorentz boosts. 
Indeed in terms of components we have for the angular velocity tensor
\begin{eqnarray}
&&\Omega_{ab}= -\frac{q}{m_E}\epsilon_{abc}B^c- \frac{ q}{m_Ev^2}(E_a v_b-E_bv_a)+ \frac{ \partial_\phi \ln A}{\gamma v^2}  \left(\partial_a\phi v_b-\partial_b\phi v_a \right)~,\nonumber\\
&&\Omega_{0a}=0~.
\end{eqnarray}
This decomposition makes the Lorentz transformations of the angular velocity more transparent as we have simply to boost $\Omega_{\mu\nu}$ and pick the ``magnetic'' part of this antisymmetric tensor to define the boosted angular velocity vector. The magnetic part $B^a$ transforms as a magnetic field whilst the ``electric'' part ${\cal E}_a$ is invariant. 

\section{A simplified derivation of the classical frequency shift}
\label{appendix:frequency}

In this appendix we present a simplified version of the derivation of the scalar field's classical contribution to the spin precession frequency that was given in Section \ref{sec:classical-precession}.  We will assume that there is no coupling between the scalar field and the fermion's angular momentum, and we drop any terms that do not contribute to the final result.  This closely follows the discussion of spin precession in \cite{Jackson:1998nia}, borrowing several standard expressions along the way.

Imagine a particle with electric charge $q$ moving with velocity $\vec v$ perpendicular to a uniform magnetic field $\vec B$. We will begin by focusing on just the electromagnetic component, ignoring any contributions from a new scalar field for now.  We will also drop terms involving $\vec E$ fields, as well as $\vec B$ fields parallel to the motion of the particle, keeping only the pieces that are non-zero in the specific case of perfect cyclotron motion.

Focusing on the case of pure electromagnetism (that is, no scalar field) for now, the relativistic equation of motion is
\begin{equation}
     m_E \frac{d \vec v}{d \tau} = q \vec v \wedge \vec B~.
\end{equation}
Since $d \tau = dt / \gamma$, we obtain
\begin{equation}
    \frac{d \vec v}{d t} = \frac{q}{\gamma m_E} \vec v \wedge \vec B~.
\end{equation}
We can rewrite this as a gyroscope equation
\begin{equation}
    \frac{d \vec v}{dt} = \vec \omega_c \wedge \vec v~,
    \label{precession-force}
\end{equation}
where we have identified the cyclotron frequency
\begin{equation}
    \vec \omega_c \equiv - \frac{q}{\gamma m_E} \vec B~.
\end{equation}
This is the angular velocity as the particle's motion traces a perfect circle.

Now we compute the rate of change of the spin $\vec s$ of the particle.  One is tempted to reach for the usual expression for a magnetic moment $\vec \mu = \frac{g q}{2 m}$ in a magnetic field\footnote{The vector $\vec s$ here is the spin vector in the rest frame of the particle.}
\begin{equation}
    \frac{d \vec s}{d t} = \vec \mu \wedge \vec B~.
\end{equation}
The issue with this is that it is not a relativistic expression, so it must be evaluated in the rest frame of the particle.  A related problem is that the rest frame of the particle is non-inertial, as it rotates with the particle.  In other words,
\begin{equation}
    \left( \frac{d \vec s}{ d t'}\right )_\mathrm{rot} = \vec \mu \wedge \vec B'~,
\end{equation}
where a prime indicates the rest frame of the particle.  We need to write this in terms of laboratory-frame quantities.  The $\vec B$ field is entirely perpendicular to the motion so we have the usual Lorentz boost
\begin{equation}
    \vec B' = \gamma \vec B~.
\end{equation}
(Again, we are ignoring any $\vec E$ fields as they do not contribute to the final result.)
Next, we convert from the particle's rest frame time $t'$ to the laboratory frame time $t$:
\begin{equation}
    \left( \frac{d \vec s}{ d t'}\right )_\mathrm{rot} = \frac{d t}{d t '} \left( \frac{d \vec s}{ d t}\right )_\mathrm{rot} = \gamma \left( \frac{d \vec s}{ d t}\right )_\mathrm{rot}~,
\end{equation}
so that ultimately we have
\begin{equation}
    \left( \frac{d \vec s}{ d t}\right )_\mathrm{rot} = \vec \mu \wedge \vec B~.
\end{equation}
Now we need to translate from a rotating frame to the inertial laboratory frame.  We use \cite{Jackson:1998nia}
\begin{equation}
    \left( \frac{d \vec s}{ d t}\right )_\mathrm{nonrot} = \left( \frac{d \vec s}{ d t}\right )_\mathrm{rot} + \vec \omega_\mathrm{T} \wedge \vec s~,
\end{equation}
which is true of any vector when translating from a rotating frame to an inertial one, where $\vec \omega_\mathrm{T}$ is the Thomas precession frequency and will be given shortly.  Summarizing, we have
\begin{equation}
    \left( \frac{d \vec s}{d t} \right)_\mathrm{nonrot} =  \left(- 
    \frac{g q}{2 m_E } \vec B + \vec \omega_\mathrm{T} \right) \wedge \vec s~.
    \label{prec-nonrot}
\end{equation}
The first term results from the usual Larmor precession while the second term is from Thomas precession, a relativistic correction for the non-rotating frame, given by
\begin{equation}
    \vec \omega_\mathrm{T} = \frac{\gamma^2}{\gamma + 1} \vec a \wedge \vec v~.
\end{equation}
This depends on the acceleration $\vec a$ of the particle in the (non-rotating) laboratory frame.  Using Eq.~\eqref{precession-force}, we can simplify this as
\begin{equation}
    \vec \omega_\mathrm{T} = - \frac{\gamma^2}{\gamma + 1} \vec v^2 \vec\omega_c = (1 - \gamma) \vec\omega_c~,
\end{equation}
where the first equality assumed that the acceleration is orthogonal to the motion of the particle, and the second equality used the identity $\vec v^2 = (\gamma^2 - 1) / \gamma^2$~.

We now identify the spin precession frequency from Eq.~\eqref{prec-nonrot}
\begin{align} \nonumber
    \vec \omega_s &= - \frac{g q}{2 m_E} \vec B + \vec \omega_\mathrm{T}~, \\
    &= - \frac{g q}{2 m_E} \vec B + (1 - \gamma) \vec \omega_c~,
\end{align}

The difference between the spin and cyclotron precession frequencies is the anomalous precession frequency
\begin{align} \nonumber
    \vec \omega_a &= \vec \omega_s - \vec \omega_c~, \\ 
    &= - \frac{g q}{2 m_E} \vec B - \gamma \vec \omega_c~.
    \label{prec-anomalous}
\end{align}
Plugging in the cyclotron frequency $\vec \omega_c$, we find
\begin{equation}
    \vec \omega_a = - \frac{q}{m_E} \left( \frac{g - 2}{2} \right) \vec B~,
\end{equation}
which is exactly correct for the idealised case of perfect cyclotron motion considered here.

Now we can examine the effects of a scalar field gradient $\vec \nabla \phi$ that is perpendicular to $\vec v$.  Using Eq.~\eqref{cyclotron-frequency} for the cyclotron frequency,
\begin{equation}
    \vec \omega_c = - \frac{q}{ \gamma m_E} \vec B - \frac{ \partial_\phi \ln A}{\gamma^2 - 1} \vec v \wedge \vec \nabla \phi~.
\end{equation}
Substituting this in to Eq.~\eqref{prec-anomalous} we find
\begin{equation}
    \vec \omega_a = - \frac{q}{m_E} \left( \frac{g - 2}{2} \right) \vec B + \frac{\gamma}{\gamma^2 - 1} (\partial_\phi \ln A) \vec v \wedge \vec \nabla \phi~,
\end{equation}
which agrees with Eq.~\eqref{BMT}, apart from the terms that were intentionally dropped for simplicity.  The remaining terms can be constructed in a manner analogous to the one presented here.

\section{Dark energy vs. modified gravity}
\label{app:DE-vs-MG}

There are several different criteria which could qualify a chameleon or a symmetron model as having something to do with dark energy.
It is well known that chameleon and symmetron models cannot act as dynamical dark energy alone on cosmological scales~\cite{Brax:2011aw, Wang:2012kj}, but it is still worth asking whether the theory can behave like dark energy on laboratory  scales. One proviso is that in the chameleon case the quantum corrections are intrinsically field dependent and therefore can alter the classical results significantly. For symmetrons, the situation is simpler and 
without satisfying at least one of the options listed here, the theory is better described as modified gravity, as it works on scales very different than those of dark energy, Einstein gravity, and cosmological expansion. Although a full examination of the cosmological consequences of the symmetron models indicated in Fig.~\ref{fig:symmetron} would be interesting, it is beyond the scope of the present work so we focus mostly on whether these modified gravity models could be associated with dark energy. A brief sketch of some of the early Universe consequences of the symmetrons with parameters as in Fig.~\ref{fig:symmetron} is given below for completeness.

\subsection{Early Universe}
The physics of symmetrons in the Universe depends on the symmetry breaking energy density $\rho_{\rm B}=\mu^2  M^2$. As we consider models where $\mu \gtrsim 10^{-6}$ GeV and $M\gtrsim 10^{-1}$ GeV, see Fig.~\ref{fig:symmetron}, the symmetry breaking energy density is $\rho_{\rm B} \gtrsim 10^{-14}\ {\rm GeV}^4$.  This energy density is always much larger than the matter density at Big Bang Nucleosynthesis $\rho_{\rm BBN}\sim 10^{-20}\ {\rm GeV}^4$ at a redshift of $z\sim 10^9$.  As a result the symmetron field essentially sits at the effective minimum of the matter-dependent potential where $\langle \phi\rangle =\frac{\mu}{\sqrt \lambda}$ with a mass $m_\phi=\sqrt 2 \mu$. For large values of $1~ {\rm MeV} \lesssim \mu \lesssim 1 ~{\rm GeV}$, the symmetrons are sufficiently heavy during BBN that they play hardly any role in the early Universe. For instance, they are always non-relativistic and as $\langle \phi\rangle$ is constant the constraint that fermion masses do not vary in the Einstein frame since the time of BBN on the variation of masses is trivially satisfied. When $1 ~{\rm keV} \lesssim \mu \lesssim 1 ~{\rm MeV}$ symmetrons could appear as a new relativistic species during BBN if they thermalise with electrons \cite{Nollett:2011aa}. As the symmetrons have a $\frac{m_e}{M^2} \phi^2 \bar e e $ coupling to electrons, the scalar cross section behaves like $\sigma \simeq \frac{m_e^2}{M^4}$ and the symmetrons are coupled to the primordial plasma for $T\gtrsim T_\phi= \frac{M^4}{m^2_e M_{\rm Pl}}$. This is larger than $1$ MeV for $M\gtrsim 10^{3}$ GeV. As a result, the lowest range of the symmetron masses would affect BBN and could be ruled out cosmologically.

\subsection{Mass}
One possible criterion is that the mass of the symmetron particle is at the dark energy scale, implying $\mu \approx \Lambda_\mathrm{DE} \equiv 2.4~\mathrm{meV}$.  One upside of this choice is that at one loop the Coleman-Weinberg correction to the potential goes as (see, e.g. \cite{Upadhye:2012vh}) $\Delta V \sim \mu^4$, generating a dark-energy-like density.  The viable symmetron parameters we found to explain the $g-2$ discrepancy all have mass $\mu > 1~ \mathrm{eV}$, so this option is ruled out.
For the chameleon with only a matter coupling, this scenario would indicate a matter coupling $M \approx 10^8~\mathrm{eV}$, which is in tension with collider and electron $g - 2$ bounds.

\subsection{Energy density}
Another possibility is for the energy offset between the symmetron's false and true vacua to account for the cosmological constant, resulting in a term in the Lagrangian of order $\sim \Lambda_\mathrm{DE}^4$.
The symmetron vev is $v = \mu / \sqrt{\lambda}$.  In a pure vacuum where $\rho_m = 0$, then the difference in energy density between the false and true vacua is
\begin{align} \nonumber
    \Delta V_\mathrm{eff} &= \frac{1}{2} \left( \frac{\rho}{M^2} - \mu^2 \right) \phi^2 + \frac{1}{4} \lambda \phi^4~, \\
    &= - \frac{\mu^4}{4 \lambda}~.
\end{align}
If we wish for this to be of order the dark energy density $\Lambda_\mathrm{DE}^4$ where $\Lambda_\mathrm{DE} = \mathrm{meV}$, then we need
\begin{equation}
    \lambda \approx \frac{\mu^4}{\Lambda_\mathrm{DE}^4} = \left( \frac{\mu}{\Lambda_\mathrm{DE}} \right)^4~.
\end{equation}
We also require $\lambda < 1$ for perturbation theory to work, so we are restricted to $\mu < \mathrm{meV}$.  All of the symmetron masses we propose to resolve the $g-2$ tension are larger than this.

For chameleons, the analogous requirement would be for the energy in the chameleon potential to match the energy density of dark energy: $V(\phi) \approx \Lambda_\mathrm{DE}^4$.  The chameleon models that resolve the $g - 2$ tension, shown in Fig.~\ref{fig:chameleon}, have $M \approx 10^{-16} \Mpl$.  This requirement leads to $\Lambda \approx \mathrm{eV}$, which is ruled out by atom interferometry.

\subsection{Critical density}
A third option, unique to the symmetron, is for the symmetron critical density to match the critical density of the universe:
\begin{equation}
    \mu^2 M^2 = 3 \Mpl^2 H_0^2~,
\end{equation}
such that the field becomes tachyonic at the present day and correspondingly mediates a very long-ranged fifth force.
Recall that for there to be any scalar effect at all, classical or quantum, we need the symmetron to exist in the symmetry-broken phase.  That required $\rho_\mathrm{vac} \ll \mu^2 M^2$.  But $\rho_\mathrm{vac} \approx 10^{-12} ~\mathrm{g / cm}^3$, which is much larger than the critical density of the universe ($10^{-29}~\mathrm{g / cm}^3$) so it is impossible for these densities to match in realistic laboratory experiments. 

\subsection{Fifth force strength}
Another option is for the strength of the unscreened fifth force to be comparable to ordinary gravity.  The coupling strength in the symmetry broken phase is $\phi / M^2 = \mu / (\sqrt{\lambda} M^4)$.  For this coupling strength to match that of gravity, we need
\begin{equation}
    \frac{\mu}{\sqrt{\lambda} M^2} \approx \frac{1}{\Mpl}~,
\end{equation}
or
\begin{equation}
    \lambda \approx \frac{\Mpl^2 \mu^2 }{M^4} = (10^{54}~\mathrm{eV}^2) \frac{\mu^2}{M^4}~.
\end{equation}
However, the symmetron parameters that resolve the $g-2$ tension are given by Eq.~\eqref{symm-qm-only}
\begin{equation}
    \lambda = \frac{m_\mu^2 \mu^2}{\delta a_\mu M^4} = (10^{24} ~\mathrm{eV}^2) \frac{\mu^2}{M^4}~.
\end{equation}
So the unscreened force in the symmetron models we uncover is much stronger than gravity. For the chameleon, the same requirement would simply set $M \approx \Mpl$, which is considerably removed from the chameleon models considered in this paper.

\subsection{Cosmological equation of state}

As well known on large cosmological scales, the chameleon tracks the minimum of the effective potential if its mass is much larger than the Hubble rate $m\gg H$. When this is the case the chameleon's energy density (for $n=1$) varies in time as a function of the scale factor as  $\rho_\phi(a) \propto a^{-3/2}$, coming from $\phi(a)\propto a^{3/2}$,  therefore implying  an equation of state $\omega_\phi=-1/2$ which is excluded observationally. Hence classical chameleons cannot lead to self-acceleration and require the addition of a cosmological constant to generate the late time acceleration of the Universe. For symmetrons, after tuning the vacuum energy to zero and assuming $\mu \gg H_0$, the cosmological energy density at the minimum of the effective potential is $\rho_\phi\simeq \mu^2 \rho/2M^2 \lambda \propto  a^{-3}$ which does not redshift like dark energy. 

\renewcommand{\em}{}
\addcontentsline{toc}{section}{References}

\bibliography{main}

\end{document}